\documentclass[aps,pre,epsf,superscriptaddress,amsmath,amssymb,amsfonts,twocolumn,showpacs]{revtex4-1}

\usepackage{graphicx}
\usepackage{epsfig}
\usepackage{dcolumn}
\usepackage{bm}
\usepackage{braket}
\usepackage{amsmath}
\usepackage{mathtools}
\usepackage{color}

\begin{document}
\title{Correlated Tunneling Dynamics of an Ultracold Fermi-Fermi\\ Mixture Confined in a Double-Well}

\author{J. Erdmann}
\affiliation{Zentrum f\"{u}r Optische Quantentechnologien,
Universit\"{a}t Hamburg, Luruper Chaussee 149, 22761 Hamburg,
Germany} 
\author{S. I. Mistakidis}
\affiliation{Zentrum f\"{u}r Optische Quantentechnologien,
Universit\"{a}t Hamburg, Luruper Chaussee 149, 22761 Hamburg,
Germany}
\author{P. Schmelcher}
\affiliation{Zentrum f\"{u}r Optische Quantentechnologien,
Universit\"{a}t Hamburg, Luruper Chaussee 149, 22761 Hamburg,
Germany} \affiliation{The Hamburg Centre for Ultrafast Imaging,
Universit\"{a}t Hamburg, Luruper Chaussee 149, 22761 Hamburg,
Germany}

\date{\today}

\begin{abstract} 

We unravel the correlated tunneling dynamics of a mass imbalanced few-body fermi-fermi mixture upon quenching the tilt of a double-well. 
The non-equilibrium dynamics of both species changes from Rabi-oscillations close to the non-interacting limit 
to a delayed tunneling dynamics for moderate interspecies repulsions. 
Considering strong interspecies interactions the lighter species experiences quantum self-trapping due to the heavier species which 
acts as an effective material barrier, while performing almost perfect Rabi-oscillations. 
The degree of entanglement, inherent in the system, is analyzed and found to be significant both at moderate and strong repulsions. 
To relate our findings with possible experimental realizations we simulate in-situ single-shot measurements and discuss how a sampling 
of such images dictates the observed dynamics. 
Finally, the dependence of the tunneling behavior on the mass ratio, the particle number in each species and the height 
of the barrier of the double-well is showcased. 
 
\end{abstract}

\maketitle

\section{Introduction}

Degenerate quantum gases in external traps offer an extraordinary level of control that enables the 
investigation of a multitude of many-body quantum phenomena \cite{bloch2008many}. 
A variety of parameters can nowadays be tuned experimentally, e.g. the particle number \cite{serwane2011deterministic,zurn2012fermionization,wenz2013few}, the 
interaction strength via Feshbach resonances \cite{inouye1998observation,chin2010feshbach} and 
the dimensionality of the confinement \cite{greiner2002quantum}. 
Beyond the realization of ultracold single component bosonic \cite{spielman2007mott, folling2007direct, trotzky2008time} or 
fermionic ensembles \cite{zwierlein2004condensation, partridge2006deformation, shin2006observation, inguscio2007gas, liao2011metastability} 
also the preparation of Bose-Bose \cite{catani2008degenerate, thalhammer2008double, petrov2015quantum}, 
Fermi-Fermi (FF) \cite{wille2008exploring,kohl2005fermionic,hadzibabic2002two} or 
Bose-Fermi mixtures \cite{stan2004observation, ospelkaus2006tuning, ahufinger2005disordered} have been achieved. 
Multicomponent fermionic systems consisting of different isotopes such as $\prescript{40}{}{K}$ \cite{wu2012ultracold, wille2008exploring}, 
$\prescript{6}{}{Li}$ \cite{wille2008exploring, moerdijk1995resonances}, $\prescript{84}{}{Rb}$ \cite{crane2000trapping} or 
$\prescript{87}{}{Sr}$ \cite{takamoto2006improved} have attracted considerable attention. 
They reveal intriguing features including superfluidity \cite{chen2005bcs, chin2006evidence}, 
quantum magnetism \cite{sowinski2018ground, hung2011exotic,yannouleas2016ultracold,koutentakis2018probing},  
insulating phases \cite{lisandrini2017topological, nataf2016chiral, zhou2016mott}, phase separation processes \cite{shin2006observation,partridge2006pairing,shin2008phase} 
and polaronic quasiparticles \cite {massignan2014polarons,scazza2017repulsive,schmidt2018universal,mistakidis2018repulsive}. 

Regarding the non-equilibrium quantum dynamics, atoms trapped in a double-well potential constitute a prototype system to study the 
correlated tunneling dynamics in a controllable manner. 
For bosons such a system represents the analog of the well-known superconducting Josephson 
junction \cite{albiez2005direct, zollner2008few, dounas2007ultracold, salgueiro2007quantum}. 
The bosonic Josephson junction exhibits various experimentally observed \cite{albiez2005direct} intriguing phenomena, 
such as Josephson oscillations, $\pi$ modes, macroscopic quantum 
self-trapping \cite{bloch2005ultracold,smerzi1997quantum,raghavan1999coherent,milburn1997quantum,sakmann2014universality, sakmann2010quantum, sakmann2011exact} 
and correlated pair tunneling \cite{chatterjee2010few,zollner2008tunneling,zollner2008few,zollner2006correlations,zollner2006ultracold,ishmukhamedov2018tunneling}. 
An extension is provided by multicomponent bosonic setups trapped 
in a double-well evincing for instance coherent quantum tunneling and self-trapping \cite{kuang2000macroscopic}, 
collapse and revival of population dynamics \cite{sun2009dynamics,naddeo2010quantum}, symmetry breaking and 
restoring scenarios \cite{satija2009symmetry} and counterflow superfluidity \cite{hu2011detecting}. 
Turning to the fermionic tunneling properties, the spin-polarized fermionic gas in a double-well has also been intensively studied mainly in three-dimensions  
unveiling coherent Josephson dynamics of two superfluid samples \cite{valtolina2015josephson,spuntarelli2007josephson}, temperature dependent deformations of 
the density profile \cite{salasnich2010quantum} and incoherent single-particle oscillations \cite{macri2013tunneling}. 
However, the tunneling dynamics of multicomponent fermionic systems has received much less 
attention \cite{sowinski2016diffusion,tylutki2017coherent,harshman2017infinite,paraoanu2002josephson,burchianti2018connecting}. 
This includes the investigation of the dynamical properties of few atom FF mixtures confined in a double-well potential 
in relation to the few-body eigenspectrum \cite{sowinski2016diffusion}, the necessity of the renormalization of the tunneling frequency in 
few Fermi polaron systems \cite{tylutki2017coherent} and the observation of Josephson oscillations in 
three-dimensional setups \cite{paraoanu2002josephson,burchianti2018connecting}. 
Besides the above-mentioned studies a systematic study of the FF mixture tunneling dynamics in a one-dimensional double-well 
covering the weak to strong interspecies correlation regimes has not been reported. 
In this setup, it would be particularly interesting to examine how the interspecies correlations modify the tunneling properties 
of the mixture i.e. the Josephson-like oscillations, quantum self-trapping or a more complex motion. 
Since the fermions of each species are spin-polarized the question of induced intraspecies correlations occurs. 
Another intriguing prospect is to analyze the interplay between the two fermionic clouds during the evolution and to examine whether 
one species can act as an effective material barrier for the other one, a result that is already known for bosons \cite{pflanzer2009material,pflanzer2010interspecies}.  
To track the non-equilibrium dynamics of the FF mixture we utilize the Multi-Layer Multi-Configurational Time-Dependent Hartree 
Method for Atomic Mixtures (ML-MCTDHX) \cite{ML-MCTDHX}, being a variational method that enables us to capture all the 
important particle correlations. 

Motivated by the recent few fermion experiments \cite{serwane2011deterministic,zurn2012fermionization,wenz2013few} we investigate the correlated 
tunneling dynamics of a FF mixture upon quenching an initially tilted double-well 
to a symmetric one, thus favoring the tunneling of both components to the other well once the relative energy offset vanishes. 
Inspecting the single- and two-particle probabilities for the fermions of each species to reside in a certain well 
we unveil the occurrence of three distinct dynamical tunneling regimes with respect to the interspecies repulsion. 
In particular, close to the non-interacting limit both components perform almost perfect Rabi-oscillations \cite{sowinski2016diffusion} with the heavier species 
exhibiting either single- or two-particle tunneling and the lighter one three-particle transport. 
At intermediate interactions the tunneling behavior of both species is significantly altered. 
The dynamics of the heavier component is characterized by a higher-order quantum superposition while 
the lighter species undergoes single-particle tunneling. 
For strong interactions the lighter species experiences quantum self-trapping due to the heavier species 
which acts as a material barrier \cite{pflanzer2009material,pflanzer2010interspecies}. 
The heavier component exhibits almost perfect Rabi-oscillations performing either three-particle or pair tunneling. 
The degree of both inter- and intraspecies correlations is found to be overall significant especially for intermediate interactions. 
To provide possible experimental evidences of the observed dynamics we simulate in-situ single-shot measurements and showcase 
that utilizing a sampling of such images the fermionic tunneling behavior can be retrieved. 
Finally, the dependence of the tunneling dynamics on the mass ratio of the two components, the particle number 
and the height of the barrier of the double-well is discussed providing ways to further control it. 

This work is structured as follows. 
In Section \ref{sec:theory} we introduce our setup and many-body treatment as well as the relevant observables.  
Our results on the FF tunneling dynamics in a double-well are presented in Section \ref{quench}. 
Section \ref{shots} provides the simulation of in-situ single-shot images. 
In Section \ref{control} we discuss the robustness of the tunneling behavior for different system parameters. 
We provide a summary of our findings and an outlook in Sec. \ref{conclusions}. 
In Appendix \ref{sec:tilt_magnitude}, the influence of the quench strength on the tunneling dynamics is briefly discussed.  
Finally, in Appendix \ref{single_shots} we provide details on the numerical implementation of the single-shot procedure, 
while in Appendix \ref{sec:numerics} we address the convergence of our many-body simulations.

\section{Theoretical Framework}\label{sec:theory}

\subsection{Setup}

We consider a FF mixture consisting of $N_A$ and $N_B$ spin polarized fermions and mass ratio $M_B=6M_A$ for the species $A$ and $B$ respectively. 
This mass-imbalanced system can be experimentally realized by considering e.g. a mixture of isotopes of 
$\prescript{40}{19}{K}$ and $\prescript{6}{3}{Li}$ \cite{wille2008exploring}. 
The mixture is confined in an one-dimensional tilted double-well external potential \cite{kierig2008single,cheinet2008counting} 
which is comprised by a harmonic oscillator possessing a frequency $\omega$ and 
a centered Gaussian with height $V_0$ and width $w$ as well as a linear tilt with tilt parameter $d$. 
Since the two species correspond to different atomic elements they possess distinct polarizations. 
This means that, experimentally using optical trapping \cite{grimm2000optical,cetina2016ultrafast}, they experience different double-well potentials. 
For convenience here we choose the frequencies of the imposed oscillators to be the same for both 
species and therefore the two fermionic components experience different double-wells 
due to their different masses, see Fig. \ref{abb:sketch} (a) and the discussion below. 
The corresponding many-body Hamiltonian reads 
\begin{align}
\mathcal{H}&=\sum\limits_{\sigma=A,B} \sum\limits_{i=1}^{N_\sigma}\left[  -\frac{\hbar^2}{2M_\sigma}\left( \frac{d}{d x_i^\sigma}\right)^2
+\frac{1}{2}M_\sigma\omega_\sigma^2(x_i^\sigma)^2 \right. \nonumber\\
 &+ \left.\frac{V_0}{w \sqrt{2\pi}} e^{-\frac{(x_i^\sigma)^2 }{2w^2}} +  d \cdot x_i^\sigma\right]\nonumber \\  &+ \sum\limits_{i=1}^{N_A} \sum \limits_{j=1}^{N_B}g_{AB}\delta(x_i^{A}-x_j^{B}).\label{Hamilt} 
\end{align} 
Operating in the ultracold regime $s$-wave scattering is the dominant interaction process and therefore the interspecies interactions can be adequately modeled by contact interactions 
that scale with the effective one-dimensional coupling strength $g_{AB}$ for the distinct fermionic species. 
Note here that since $s$-wave scattering is forbidden for spin-polarized fermions due to the antisymmetry of the fermionic 
wavefunction \cite{pethick2002bose, lewenstein2012ultracold}, the fermions of the same species are considered to be non-interacting 
and thus only interspecies interactions are involved. 
The effective interspecies one-dimensional coupling strength \cite{olshanii1998atomic} is given by  
${g_{AB}} =\frac{{2{\hbar ^2}{a^s_{AB}}}}{{\mu a_ \bot ^2}}{\left( {1 - {\left|{\zeta (1/2)} \right|{a^s_{AB}}}/{{\sqrt 2 {a_ \bot }}}} \right)^{ -
1}}$, where $\zeta$ is the Riemann zeta function and $\mu=\frac{M_AM_B}{M_A+M_B}$ denotes the corresponding reduced mass. 
The transversal length scale is ${a_ \bot } = \sqrt{\hbar /{\mu{\omega _ \bot }}}$, where ${{\omega _ \bot }}$ is the 
frequency of the transversal confinement, and ${a^s_{AB}}$ refers to the three-dimensional $s$-wave
scattering length between the two species. 
Experimentally $g_{AB}$ is tunable either by ${a^s_{AB}}$ via  Feshbach resonances \cite{kohler2006production, chin2010feshbach} or by ${{\omega _ \bot }}$ 
and the resulting confinement-induced resonances \cite{olshanii1998atomic,kim2006suppression}. 

In the following we shall rescale our Hamiltonian in units of $\hbar  \omega_{\perp}$. 
Then, the corresponding length, time, and interaction strength scales are given in units of
$\sqrt{\frac{\hbar}{M_{A} \omega_{\perp}}}$, $\omega_{\perp}^{-1}$ and $\sqrt{\frac{\hbar^3 \omega_{\perp}}{M_{A}}}$ respectively. 
Accordingly, the amplitude of the Gaussian barrier $V_0$, its width $w$, the tilt parameter $d$ and the frequency of the harmonic oscillator $\omega$ 
are expressed in terms of $\sqrt{ \frac{ \hbar^3 \omega_{\perp} }{ M_{A} }}$, $\sqrt{\frac{\hbar}{M_{A} \omega_{\perp}}}$, 
$\sqrt{M_{A} \omega_{\perp}^3\hbar^3}$ and $\omega_{\perp}$. 
Finally, in order to limit the spatial extension of our system we impose hard-wall boundary conditions at $x_\pm=\pm40$. 

Our system is initially prepared in the many-body ground state of the tilted double-well with a complete population imbalance, namely all 
fermions of both species reside in their corresponding left well ($-40<x<0$), see also Fig. \ref{abb:sketch} (a). 
Note that throughout this work, unless it is stated otherwise, we use $V_0=1$ and $w=0.1$ thus having two doublets below the maximum of the barrier. 
The left well is, of course, significantly energetically favorable due to the presence of the linear external tilt $dx$ for $d$ 
sufficiently large. 
As an illustration Fig. \ref{abb:sketch_1} shows the ground state population of the left well of each species [see also Eq. (\ref{one_body_prob})] for varying tilt magnitude $d$ 
for the cases $N_{\sigma}=5$ ($N_A=N_B$) when $g_{AB}=0.1$ and for $N_{\sigma}=3$ with $g_{AB}=0.1$ and $g_{AB}=4.0$.  
We observe that the population imbalance depends on both $N_{\sigma}$ and $g_{AB}$. 
However, for $d=0.2$ each species is fully localized in its left well independently of the interspecies 
interaction and the particle number. 
This independence of $g_{AB}$ is caused by the fact that for $d=0.2$ the distinct fermionic clouds are 
non-overlaping, see also Fig. \ref{abb:sketch} (a). 
To ensure that both species reside initially in their left well we therefore use $d=0.2$. 
To induce the dynamics in the symmetric double-well we quench at $t=0$ the asymmetry to $d=0$ and let the system evolve in time. 
Quenching the tilt to zero favors the tunneling of both components to the right well as the corresponding energy offset 
between the two distinct wells vanishes, see also Fig. \ref{abb:sketch} (b). 
It is worth mentioning that the case of different harmonic oscillator frequencies for each species does not yield 
fundamentally different tunneling phenomena but rather results in distinct tunneling frequencies of the two species and renders 
the tunneling regions to be presented below wider with respect to the corresponding interspecies interaction. 
We have checked this for the experimentally relevant \cite{cetina2016ultrafast} values $\omega_B=0.6\omega_A$ and $M_B=(40/7)M_A$ (results not shown here for brevity). 
Note that the tunneling dynamics can also be induced for a final asymmetry, i.e. finite values of $d$, where the left well 
is energetically favorable (see Appendix A).

\subsection{Many-Body Wavefunction Ansatz} 

To examine the quench induced tunneling dynamics of the FF mixture within the double-well 
we resort to ML-MCTDHX \cite{ML-MCTDHX}. 
Within this approach, the many-body wavefunction is expanded with respect to a time-dependent and variationally 
optimized basis, allowing us to take into account both the inter and the intraspecies correlations. 
To incorporate inter and intraspecies correlations $M$ distinct species functions, 
$\Psi^{\sigma}_k (\vec x^{\sigma};t)$ with $\vec x^{\sigma}=\left( x^{\sigma}_1, \dots, x^{\sigma}_{N_{\sigma}} \right)$ being the spatial $\sigma=A,B$ species coordinates, 
for each component consisting of $N_{\sigma}$ fermions are firstly introduced. 
It holds that $M\le \min(\dim(\mathcal{H}^A),\dim(\mathcal{H}^B))$ with $\mathcal{H}^{\sigma}$ being the Hilbert space of the $\sigma$-species. 
Accordingly, the many-body wavefunction $\Psi_{MB}$ is expressed as a truncated Schmidt decomposition \cite{horodecki2009quantum} of rank $M$ 
\begin{equation}
\Psi_{MB}(\vec x^A,\vec x^B;t) = \sum_{k=1}^M \sqrt{ \lambda_k(t) }~ \Psi^A_k (\vec x^A;t) \Psi^B_k (\vec x^B;t).    
\label{Eq:WF}
\end{equation} 
The Schmidt coefficients $\lambda_k(t)$ in decreasing order are referred to as the natural species populations of the $k$-th 
species function $\Psi^{\sigma}_k$ of the $\sigma$-species and provide a measure of the system's entanglement (see also Sec. \ref{observables}).  
Indeed, the system is termed entangled \cite{roncaglia2014bipartite} or interspecies correlated when at least two distinct $\lambda_k(t)$ are nonzero, 
thus preventing the total many-body state [Eq. (\ref{Eq:WF})] to be a direct product of two states. 

Moreover in order to include interparticle correlations each of 
the species functions $\Psi^{\sigma}_k (\vec x^{\sigma};t)$ is expanded using the determinants of $m^{\sigma}$ distinct 
time-dependent fermionic single-particle functions (SPFs), $\varphi_1,\dots,\varphi_{m_{\sigma}}$, namely  
\begin{equation}
\begin{split}
&\Psi_k^{\sigma}(\vec x^{\sigma};t) = \sum_{\substack{n_1,\dots,n_{m_{\sigma}} \\ \sum n_i=N}} C_{k,(n_1,
\dots,n_{m_{\sigma}})}(t)\times \\ &\sum_{i=1}^{N_{\sigma}!} {\rm sign}(\mathcal{P}_i) \mathcal{P}_i
 \left[ \prod_{j=1}^{n_1} \varphi_1(x_j;t) \cdots \prod_{j=1}^{n_{m_{\sigma}}} \varphi_{m_{\sigma}}(x_j;t) \right].  
 \label{Eq:SPFs}
 \end{split}
\end{equation} 
In the latter expression, $\mathcal{P}$ denotes the permutation operator exchanging the particle configuration within 
the SPFs and $\rm{sign}(\mathcal{P}_i)$ refers to the sign of the corresponding permutation. 
Also, $C_{k,(n_1,\dots,n_{m_{\sigma}})}(t)$ are the time-dependent expansion coefficients of a particular determinant and $n_i(t)$ 
is the occupation number of the SPF $\varphi_i(\vec{x};t)$. 
In the following, we shall define that each fermionic species possesses intraspecies correlations if more than $N_{\sigma}$ SPFs are substantially occupied, otherwise 
the state of a species reduces to the Hartree-Fock ansatz \cite{pethick2002bose,giorgini2008theory,pitaevskii2016bose}. 
Following e.g. the Dirac-Frenkel variational principle \cite{frenkel1932wave,dirac1930note} for the many-body ansatz [see Eqs.~(\ref{Eq:WF}), (\ref{Eq:SPFs})] 
yields the ML-MCTDHX equations of motion \cite{ML-MCTDHX} for a binary fermionic mixture. 
These consist of $M^2$ linear differential equations of motion for the coefficients $\lambda_i(t)$, which are coupled to a set of 
$M$[${m_A}\choose{N_A}$+$ {m_B}\choose{N_B}$] non-linear integro-differential equations for the species functions and $m^A+m^B$  
integro-differential equations for the SPFs. 
Finally, let us mention that ML-MCTDHX is able to operate within different approximation 
orders, e.g. it reduces to the Hartree-Fock equation for $M=1$ and $m^{\sigma}=N_{\sigma}$.

\subsection{Observables of Interest} \label{observables}

In this section we briefly introduce the main observables that will be subsequently employed for the 
interpretation of the tunneling dynamics. 

On the one-body level the tunneling dynamics can be examined by inspecting the population imbalance 
between the left and right wells during the time evolution. 
To this end, we measure the expectation value of the one-body density \cite{zollner2008tunneling} e.g. in the left well 
\begin{equation}
\braket{ \rho^{(1)}_\sigma(t)}_L=\int_{x < 0 }d x \, \rho^{(1)}_\sigma(x,t). \label{one_body_prob} 
\end{equation}
Here, $\rho^{(1)}_\sigma(x;t)=\langle\Psi_{MB}(t)|\hat{\Psi}^{\sigma \dagger}(x)\hat{\Psi}^\sigma(x)|\Psi_{MB}(t)\rangle$ denotes  
the $\sigma$-species one-body density, which is normalized to the corresponding particle number $N_{\sigma}$. 
$\Psi^{\sigma \dagger}(x)$ [$\Psi^{\sigma}(x)$] is the fermionic field operator that 
creates (annihilates) a $\sigma$-species fermion at position $x$. 
As it is evident from Eq. (\ref{one_body_prob}), $\braket{ \rho^{(1)}_\sigma(t)}_L$ essentially measures how many $\sigma$-species particles reside in the left well and it 
is normalized to the corresponding number of $N_{\sigma}$ particles such that 
$\braket{ \rho^{(1)}_\sigma(t)}_L=N_\sigma - \braket{\rho^{(1)}_\sigma(t)}_R$ holds. 

To unveil the two-body correlation mechanisms that are responsible for the observed tunneling dynamics we employ the pair probability 
$p^{(2)}_{\sigma \sigma ^\prime}$ \cite{zollner2008tunneling, chatterjee2013ultracold} in the course of the evolution. 
It is based on the diagonal two-body reduced density matrix 
$\rho^{(2)}_{ \sigma \sigma'}(x_1,x_2;t)=\bra{\Psi_{MB}(t)}\Psi^{\sigma \dagger}(x_1)\Psi^{\sigma' \dagger}(x_2)\Psi^{\sigma}(x_1)\\\Psi^{\sigma'}(x_2)\ket{\Psi_{MB}(t)}$ 
which refers to the probability of measuring two fermions of the same ($\sigma=\sigma'$) or different 
species ($\sigma\neq\sigma'$) located at positions $x_1$, $x_2$ at time $t$ respectively \cite{mistakidis2017correlation}. 
In our case $p^{(2)}_{\sigma \sigma ^\prime}$ provides the probability to find two fermions of the same or different species within the same well and it is given by  
\begin{align}
p^{(2)}_{\sigma \sigma ^\prime}&= \frac{1}{Z}\langle\Theta(x_1)\Theta(x_2)+\Theta(-x_1)\Theta(-x_2)\rangle \nonumber\\
&=\frac{1}{Z}\int_{x_1\cdot x_2\geq 0} d x_1 d x_2 \,\rho^{(2)}_{\sigma \sigma ^\prime}(x_1,x_2;t). \label{pair_prob} 
\end{align}
The normalization used corresponds to $Z=N_{\sigma} N_{\sigma^\prime}$ for $\sigma\neq \sigma ^\prime$ and 
$Z={N_{\sigma} (N_{\sigma}-1)}/{2}$ for $\sigma= \sigma ^\prime$. 
Alternatively, $p^{(2)}_{\sigma \sigma ^\prime}$ can be seen as a measure for the number of pairs that can be found 
in both wells and it is normalized to unity. 

To expose the degree of both inter- and intraspecies correlations during the time-evolution 
for increasing interspecies repulsions we measure the entanglement and 
the fragmentation of the FF mixture \cite{koutentakis2018probing,cao2017collective,fasshauer2016multiconfigurational}. 
In this way, also the corresponding departure from a Hartree-Fock state can be deduced. 
To quantify the presence of interspecies correlations or entanglement we calculate the eigenvalues
$\lambda_k$ of the species reduced density matrix
$\rho^{N_{\sigma}} (\vec{x}^{\sigma}, \vec{x}'^{\sigma};t)=\int d^{N_{\sigma'}} x^{\sigma'} \Psi^*_{MB}(\vec{x}^{\sigma}, 
\vec{x}^{\sigma'};t) \Psi_{MB}(\vec{x}'^{\sigma},\vec{x}^{\sigma'};t)$, where $\vec{x}^{\sigma}=(x^{\sigma}_1, \cdots, 
x^{\sigma}_{N_{\sigma-1}})$, and $\sigma\neq \sigma'$ [see also Eq. (\ref{Eq:WF})]. 
It is known that when multiple eigenvalues of $\rho^{N_{\sigma}}$ are macroscopically large the system is referred to as species entangled 
or interspecies correlated, otherwise it is said to be non-entangled. 
A well-known measure to quantify the degree of the system's entanglement is the 
Von-Neumann entropy \cite{yu2009short,catani2009entropy}, $S$, which is based on the natural 
populations $\lambda_i(t)$ of the species functions [see Eq. (\ref{Eq:WF})] 
\begin{align}
S(t)=-\sum\limits_{i=1}^M \lambda_{i}(t)\ln[\lambda_i(t)]. 
\end{align} 
Indeed within the Hartree-Fock, non-entangled, limit $S(t)=0$ since $\lambda_1(t)=1$ while for a beyond Hartree-Fock state 
where more than a single $\lambda_i$ is nonzero $S(t)\neq0$. 
\begin{figure}[t!]
	\includegraphics[width=0.49\textwidth]{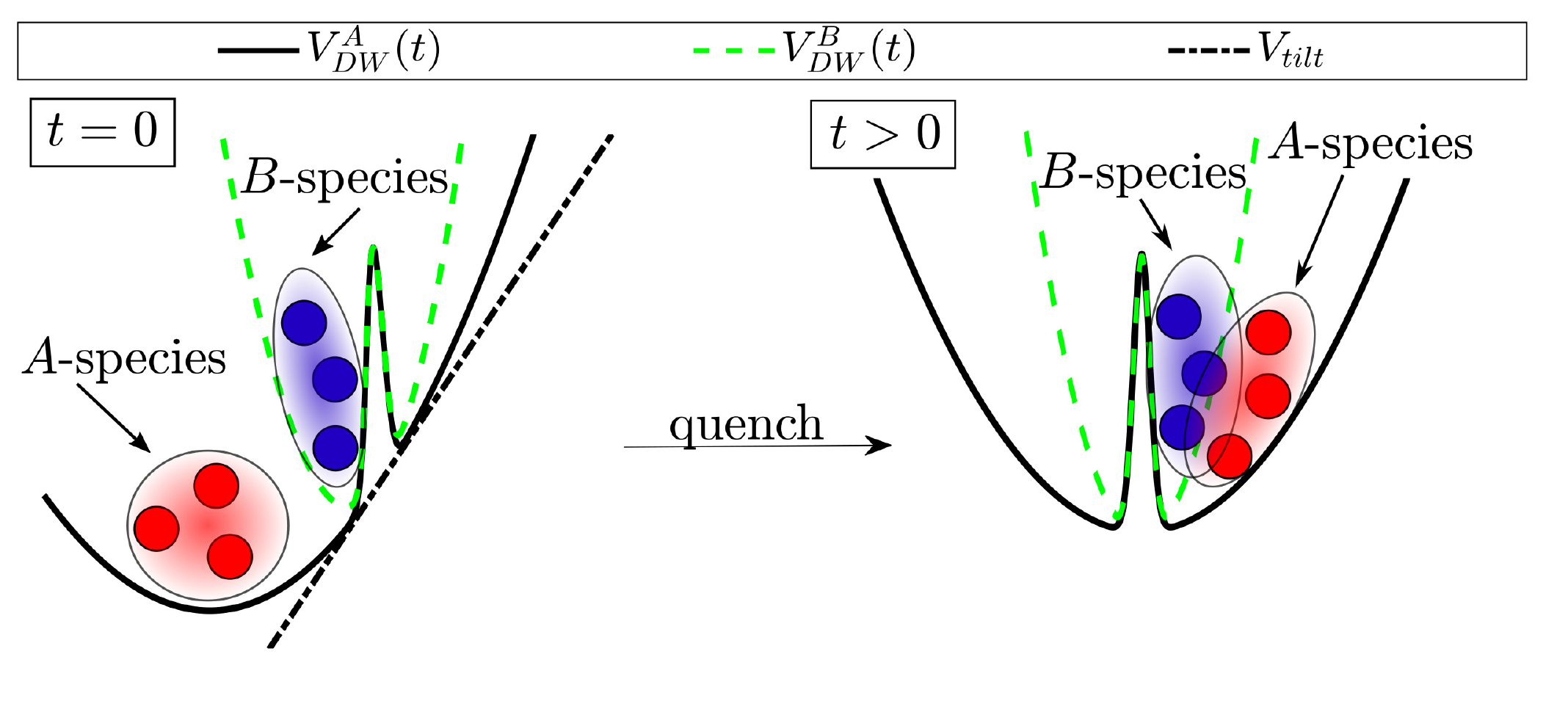}
	\caption{ Schematic representation of the quench protocol consisting of (a) an initially tilted double-well with $V_{tilt}=dx>0$ to (b) a 
	symmetric one with $V_{tilt}=d=0$. 
	The quench leads to the tunneling dynamics of the two fermionic species being initially localized in the left well of their corresponding 
	double-well. 
	In the sketch the double-wells $V_{DW}^A$, $V_{DW}^B$ for each species are shown as well as the initial tilt $V_{tilt}$ (see legend).}
	\label{abb:sketch} 
\end{figure} 

To reveal the fragmented nature of each species we rely on the eigenvalues of the one-body reduced density matrix 
of the $\sigma$-species $\rho^{(1)}_{\sigma}(x,x';t)=\bra{\Psi_{MB}(t)}\Psi^{\sigma \dagger}(x')\Psi^{\sigma}(x)\ket{\Psi_{MB}(t)}$ \cite{streltsov2008formation}.  
The eigenfunctions of $\rho^{(1)}_{{\sigma}}(x,x')$, are the so-called $\sigma$-species natural orbitals, $\phi^{\sigma}_i(x;t)$, which 
we consider to be normalized here to their corresponding eigenvalues $n^{\sigma}_i(t)= \int d x~ \left| \phi^{\sigma}_i(x;t) \right|^2$.
It can be shown that when $\Psi_{MB}(\vec x^A,\vec x^B;t) \to \Psi_{HF}(\vec x^A,\vec x^B;t)$ 
the corresponding natural populations obey $\sum_i^{N^{\sigma}}n_i^{\sigma}(t)=N_{\sigma}$, $n_{i>N_{\sigma}}^{\sigma}(t)=0$ 
and the corresponding Hartree-Fock wavefunction is retrieved. 
Therefore, if more than $N_{\sigma}$ natural orbitals are occupied, the system is said to be fragmented and the corresponding degree of 
fragmentation can be quantified via 
\begin{equation}
F_{\sigma}(t) = N_{\sigma} - \sum\limits_{i=1}^{N_{\sigma}}n_i^\sigma(t).
\end{equation} 
This quantity serves as a theoretical measure for the occupation of the $m^{\sigma} - N_{\sigma}$ least occupied 
natural orbitals and thus for the deviation from a Hartree-Fock state when $F_{\sigma}>0$.

\section{Quench Induced Tunneling Dynamics}\label{quench}

\begin{figure}[t!]
	\includegraphics[width=0.49\textwidth]{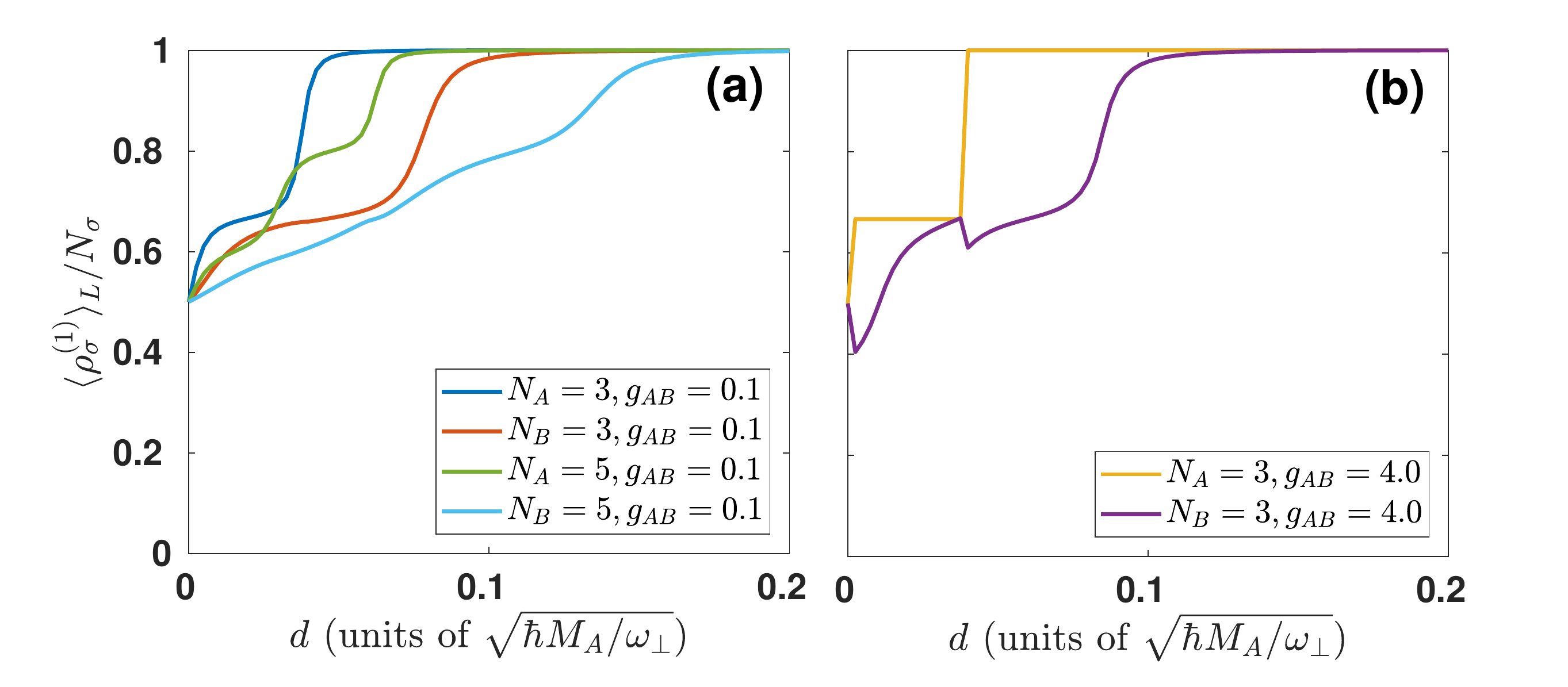}
	\caption{ Ground state population, $\braket{\rho^{(1)}_{\sigma}}_L/N_{\sigma}$ [see also Eq. (\ref{one_body_prob})], of the left well for each species for varying tilt $d$. 
	Different particle numbers $N_{\sigma}$ and interspecies repulsions $g_{AB}$ are shown (see legend). }
	\label{abb:sketch_1} 
\end{figure}

We consider a mass imbalanced repulsively interacting, $g_{AB}$, FF mixture consisting of $N_A=N_B=3$ fermions with $M_A=1$ and $M_B=6$. 
The system is initially prepared in the ground state of the tilted double-well described in Eq. (\ref{Hamilt}) with tilt parameter $d=0.2$, frequency $\omega=0.1$, barrier 
height $V_0=1$ and width $w=1$. 
For these values and since $M_B>M_A$ the heavier B-species experiences a much more localized double-well around $x=0$ when compared to the double-well 
of the lighter A-species. 
Moreover due to the tilt, each species is found to be fully localized in its respective double-well, see Fig. \ref{abb:sketch} (a) for a schematic respresentation. 
Then the distinct fermionic clouds are non-overlaping for $t=0$, since $d=0.2$ see Fig. \ref{abb:sketch_1}, and therefore phase separated at $x<0$. 
Concluding, this non-overlaping behavior between the two initial fermionic clouds is caused by their mass imbalance \cite{shin2006observation,shin2008phase} and 
renders the ground state to be essentially independent of $g_{AB}$, see Fig. \ref{abb:sketch_1}. 
To examine the tunneling dynamics of the mass imbalanced FF mixture we perform a quench of the initially 
titled double-well with $d=0.2$ to a symmetric one i.e. $d=0$, see for instance Fig. \ref{abb:sketch} (b), keeping fixed all other parameters for 
a specific $g_{AB}$ and covering the regime from weak to strong interactions namely $g_{AB}\in[0,5]$. 

\begin{figure*}[t!]
	\includegraphics[width=1\textwidth]{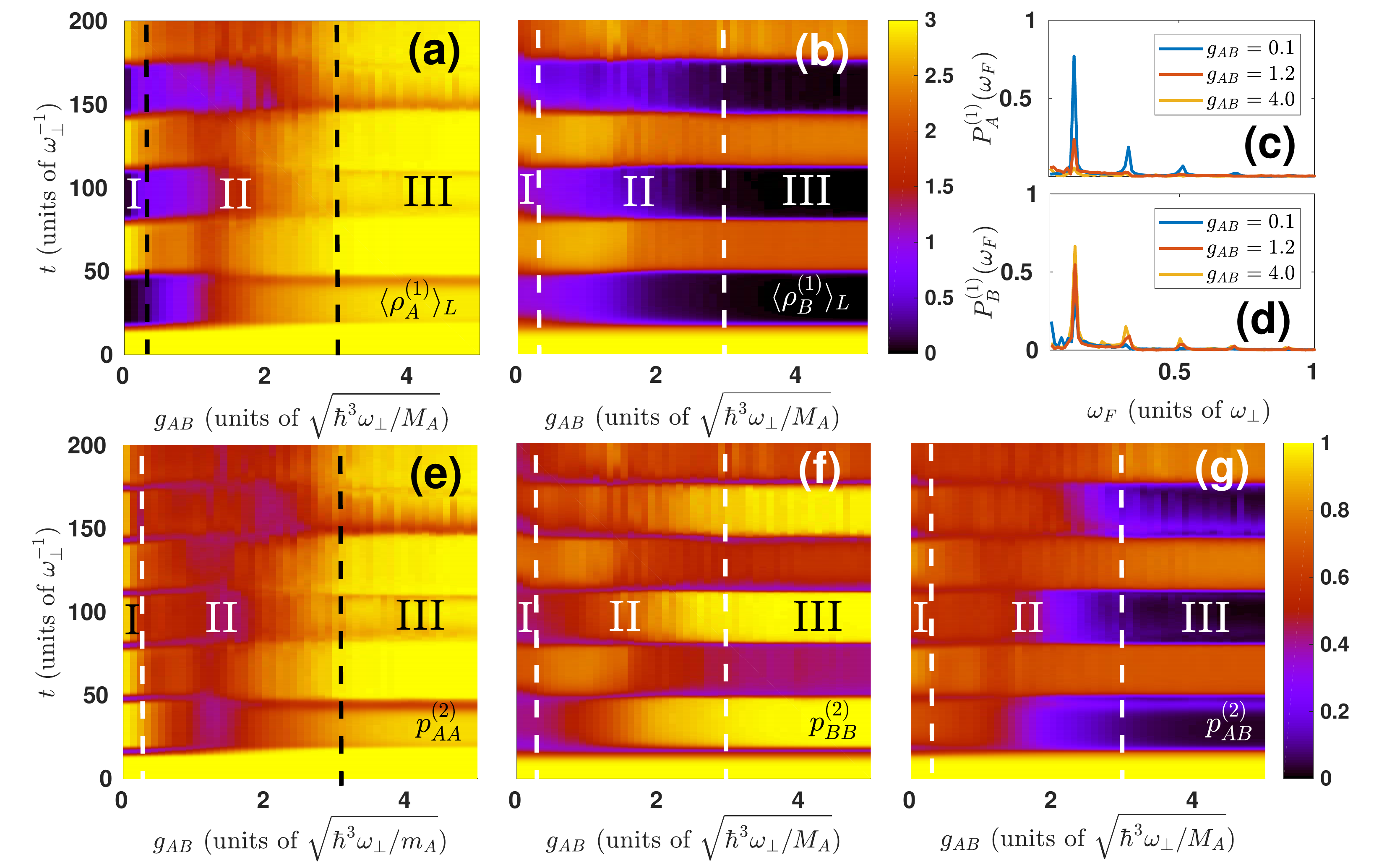}
	\caption{ Time evolution of the expectation value of the one-body density in the left well, $\braket{\rho^{(1)}_{\sigma}(t)}_L$, of the (a) $\sigma=A$ and (b) $\sigma=B$ species 
	for increasing interspecies repulsion $g_{AB}$. 
	Fourier spectrum of (c) $\braket{\rho^{(1)}_{A}(t)}_L$ and (d) $\braket{\rho^{(1)}_{B}(t)}_L$ for different interspecies repulsions (see legends).   
	Evolution of the two-body intraspecies probability, $p^{(2)}_{\sigma\sigma}(t)$, to find two fermions of the (e) $\sigma=A$ and (f) $\sigma=B$ species to reside in the 
	same (either left or right) well for varying $g_{AB}$.  
	(g) Interspecies two-body probability, $p^{(2)}_{AB}(t)$, for one fermion of each species to be within the same well for varying $g_{AB}$.  
	In all cases the system consists of $N_A=N_B=3$ fermions initially confined in a tilted double-well with $d=0.2$ and we follow the tunneling dynamics after a quench to a symmetric 
	double-well i.e. $d=0$.} 
	\label{abb:explefth1} 
\end{figure*} 

\subsection{One- and Two-Body Tunneling Probabilities}

To monitor the overall tunneling dynamics on the one-body level for different values of $g_{AB}$ we rely on the expectation value of the 
single-particle density within the left well $\braket{ \rho^{(1)}_\sigma(t)}_L$ [see also Eq. (\ref{one_body_prob})] for each species. 
$\braket{ \rho^{(1)}_\sigma(t)}_L$ essentially provides the percentage of the $\sigma$-species fermions within the left well or alternatively speaking 
the population imbalance between the two-sides of the double-well and offers therefore an adequate measure for the tunneling dynamics 
on the one-body level. 
Our primary aim is to investigate whether distinct tunneling regions for the fermions can be observed for a varying 
interspecies repulsion \cite{sowinski2016diffusion}. 
To conclude upon a certain tunneling regime we study $\braket{ \rho^{(1)}_\sigma(t)}_L$ as well as its 
spectrum, $P^{(1)}_{\sigma}(\omega)=\operatorname{Re} \{ \frac{1}{\pi}\int dt \braket{ \rho^{(1)}_\sigma(t)}_L e^{i\omega t} \}$, in order to infer about the 
corresponding participating mode frequencies. 
Figures \ref{abb:explefth1} (a) and (b) show $\braket{ \rho^{(1)}_A(t)}_L$ and $\braket{ \rho^{(1)}_B(t)}_L$ respectively for increasing $g_{AB}$. 
Inspecting $\braket{ \rho^{(1)}_\sigma(t)}_L$ with respect to $g_{AB}$ we observe that the tunneling behavior of each species can 
be divided into three different regions denoted as I, II and III in Figs. \ref{abb:explefth1} (a) and (b). 
Region I refers to weak interactions, namely $0<g_{AB}<0.2$, and the species $A$ ($B$) undergo a three (two)-mode tunneling 
motion, see also Figs. \ref{abb:explefth1} (c), (d), where almost all fermions of each species oscillate back and forth between 
their left- and right wells \cite{sowinski2016diffusion,cao2017collective}. 
Notice that $\braket{ \rho^{(1)}_\sigma(t)}_L$ takes values between 0 and 3, see also Table \ref{table}. 
For increasing repulsion, $0.2<g_{AB}<3$, we enter region II where the tunneling process is modified with respect to 
region I, especially for the $A$-species. 
In particular the tunneling oscillations of the $A$- and $B$-species are characterized by one and two distinct frequencies respectively, 
see Figs. \ref{abb:explefth1} (c) and (d). 
Regarding the tunneling behavior of the fermions this region can also be seen as a transition region between the weak and 
strong interaction regimes. 
For even stronger interactions $3<g_{AB}<5$, we realize region III where the $A$-species remains mainly in the left well, 
while the $B$-species still exhibits a strong amplitude two-frequency [see Figs. \ref{abb:explefth1} (c), (d)] tunneling dynamics. 
As we shall argue below this latter interplay between the two species is caused by the fact that in this strongly interacting regime 
the heavier $B$-species acts as a material barrier for the lighter $A$-species, thus supressing the tunneling of the latter (see also Sec. \ref{single_particle_density}). 
Note also that for the $A$-species this region III is reminiscent to the few-body analogue of quantum self-trapping exhibited for strongly 
interacting bosons trapped in a double-well \cite{chatterjee2010few,zollner2008tunneling,zollner2008few,zollner2006correlations,zollner2006ultracold}. 
We actually observe a small amplitude tunneling for times 
$t>800$ (not shown here), while the observed tunneling amplitude of the $B$-species is even higher than in region I. 

To gain deeper insights into the tunneling motion within and between the species and in order to reveal the underlying correlation mechanisms we 
invoke the pair probability $p_{\sigma\sigma\prime}^{(2)}(t)$ [Eq. (\ref{pair_prob})]. 
This quantity together with $\braket{ \rho^{(1)}_\sigma(t)}_L$ enables us to identify the dominant particle configurations 
$\ket{N_L,N_R}_{\sigma}$ occuring in the course of the dynamics.  
Here, $N_L$ ($N_R$) refers to the number of $\sigma$-species fermions in the left (right) well.  
We remark that due to the Pauli exclusion principle \cite{giorgini2008theory,pitaevskii2016bose} different fermions of the same species that reside in the 
same well populate energetically consecutive single-particle bands e.g. 
for $\ket{2,1}_A\equiv\ket{1^0\otimes1^1,1^0}$ where the upper index refers to the band number. 
The same holds for more than two fermions of different species that are in the same well. 
Referring to a certain time instant $t=t_1$ when e.g. $p_{AA}^{(2)}(t_1)\approx2/3$
and $2<\braket{ \rho^{(1)}_A(t_1)}_L<3$ the dominant $A$-species configuration corresponds to the superposition 
$\ket{2,1}_A+\ket{3,0}_A$. 
In the limiting case of $p_{AA}^{(2)}(t_1)\approx1$ and $\braket{ \rho^{(1)}_A(t_1)}_L\approx3$ the $\ket{3,0}_A$ state is mainly contributing. 
It is also worth stressing at this point that in order to systematically conclude upon certain number state configurations 
one needs to rely on a projection of the numerically obtained many-body wavefunction to a multiband Wannier number state basis as it has been 
demonstrated in \cite{mistakidis2017mode,mistakidis2014interaction,mistakidis2015negative,mistakidis2015resonant,koutentakis2017quench} e.g. for the single component bosonic case. 
However, such a numerical analysis in our FF mixture case is computationally challenging, since we need to take into account a large sample of 
number states in order to form a complete basis. 
Below we discuss the behavior of $p_{\sigma\sigma\prime}^{(2)}(t)$ within each of the above-mentioned tunneling regions I, II and III. 
\begin{figure}
	\includegraphics[width=0.5\textwidth]{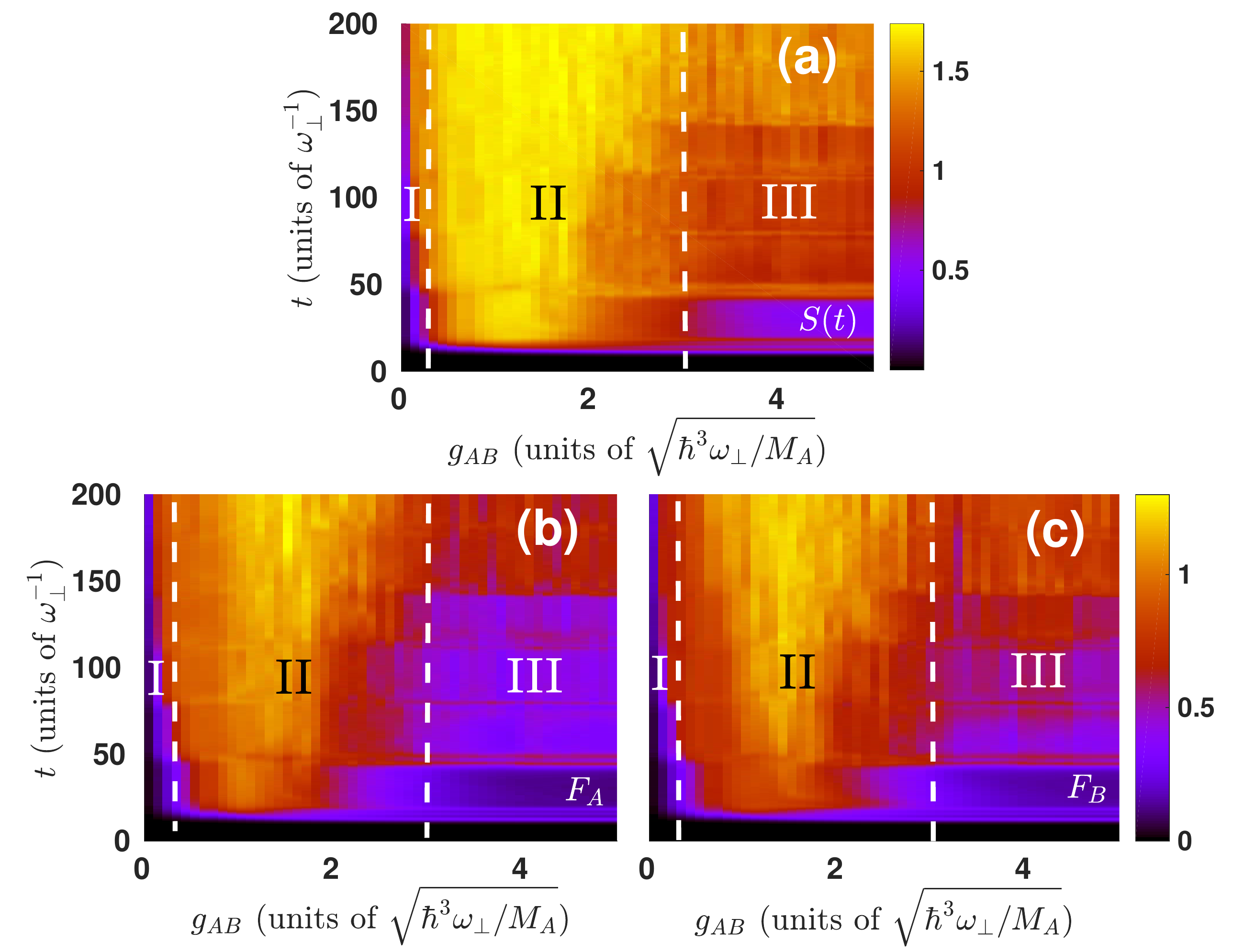}
	\caption{Time evolution of (a) the Von-Neumann entropy, $S(t)$, and the degree of fragmentation $F_{\sigma}(t)$ of 
	the (b) $\sigma=A$ and (c) $\sigma=B$ species for increasing interspecies repulsion $g_{AB}$. 
	The system consists of $N_A=N_B=3$ fermions prepared in a tilted double-well with $d=0.2$ 
	which is subsequently quenched to a symmetric double-well i.e. $d=0$.}
	\label{abb:entanglement} 
\end{figure}

 \begin{table}
  
 \begin{center}
  \begin{tabular}{|c| c| c |} 
  \hline
  Region I & Region II & Region III \\ [0.5ex] 
  \hline\hline
   $0<g_{AB}<0.2$ & $0.2<g_{AB}<3.0$ & $3.0<g_{AB}<5.0$ \\ 
  \hline
   $0<\braket{\rho^{(1)}_{A}}_L<3$ & $0.9<\braket{\rho^{(1)}_{A}}_L<2.7$ & $2.9<\braket{\rho^{(1)}_{A}}_L<3.0$ \\
  \hline
  $1.0<\braket{\rho^{(1)}_{B}}_L<2.5$ & $0.1<\braket{\rho^{(1)}_{B}}_L<2.7$ & $0<\braket{\rho^{(1)}_{b}}_L<2.1$ \\
  \hline
  $1.0<p^{(2)}_{AA}(t)<1.0$ & $0.7<p^{(2)}_{AA}(t)<0.8$ & $0.9<p^{(2)}_{AA}(t)<1.0$ \\
  \hline
  $0.4<p^{(2)}_{BB}(t)<0.7$ & $0.5<p^{(2)}_{BB}(t)<0.9$ & $0.4<p^{(2)}_{BB}(t)<1.0$ \\ 
  \hline
  $0.6<p^{(2)}_{AB}(t)<0.8$ & $0.2<p^{(2)}_{AB}(t)<0.7$ & $0<p^{(2)}_{AB}(t)<0.7$ \\
  \hline
  $0<S(t)<0.8$ & $0<S(t)<1.7$ & $0<S(t)<1.4$ \\
  \hline
  $0<F_{A}(t)<0.2$ & $0<F_{A}(t)<1.1$ & $0<F_{A}(t)<0.7$ \\  
  \hline
   $0<F_{B}(t)<0.2$ & $0<F_{B}(t)<1.2$ & $0<F_{B}(t)<0.8$ \\ [1ex]
  \hline
 
 \end{tabular}
 \caption {Summary of the different tunneling regions of the FF mixture with respect to 
 the interspecies interaction strength $g_{AB}$ and the corresponding range of the one- and two-body tunneling 
 probabilities [$\braket{\rho^{(1)}_{\sigma}(t)}_L$, $p^{(2)}_{\sigma\sigma'}(t)$], the Von-Neuman entropy $S(t)$ and 
 the degree of fragmentantion $F_{\sigma}(t)$ for each species.   }\label{table}
 \end{center}
 
 \end{table}

Regarding the $A$-species we observe that for both weak (region I) and strong interspecies interactions (region III) all three 
fermions reside in the same well since $p_{AA}^{(2)}(t)\approx1$, see Fig. \ref{abb:explefth1} (c) and also Table \ref{table}. 
The origin of the latter mechanism, in each region, can be better understood in combination with the corresponding behavior 
of $\braket{ \rho^{(1)}_A(t)}_L$ [Fig. \ref{abb:explefth1} (a)]. 
In particular, within the weak interaction regime (region I) all three particles perform a simultaneous tunneling from the left 
to the right-hand well and vice versa, since $\braket{ \rho^{(1)}_A(t)}_L$ oscillates between 0 and 3. 
Then, the system predominantly oscillates between the configurations $\ket{3,0}_A$ and $\ket{0,3}_A$. 
However, for strong interactions (region III) the $A$-species fermions remain trapped in the left well 
as $2.9<\braket{ \rho^{(1)}_A(t)}_L<3$ and the main contribution stems from the number state $\ket{3,0}_A$. 
In sharp contrast, within the intermediate interaction regime (region II) we observe that $0.7<p^{(2)}_{AA}(t)<0.8$ while 
$0.9<\braket{ \rho^{(1)}_A(t)}_L<2.7$.  
As a consequence, at least one fermion tunnels from the left to the right well 
or better it is delocalized over both wells, see also Sec. \ref{single_particle_density}. 
This indicates that the states $\ket{2,1}_A$ and $\ket{1,2}_A$ dominantly contribute. 

Next, we turn our attention to the dynamics of the $B$-species and show $p^{(2)}_{BB}(t)$ for varying 
$g_{AB}$ in Fig. \ref{abb:explefth1} (d). 
As it can be seen $p^{(2)}_{BB}(t)$ differs significantly from $p^{(2)}_{AA}(t)$, see also Table \ref{table}. 
In region I, it oscillates between $p^{(2)}_{BB}\approx1/3$ and $p^{(2)}_{BB}\approx2/3$ while 
$1.0<\braket{ \rho^{(1)}_B(t)}_L<2.5$ [Fig. \ref{abb:explefth1} (b)]. 
Thus, the $B$-species cloud mainly tunnels between the $\ket{1,2}_B$ state (e.g. at $t\approx30$) 
and the superposition $\ket{2,1}_B+\ket{3,0}_B$ (e.g. at $t\approx60$) resulting in two and single particle tunneling respectively. 
In region II, $p^{(2)}_{BB}(t)$ exhibits small amplitude fluctuations around the value 0.7, 
while $0.1<\braket{ \rho^{(1)}_B(t)}_L<2.7$. 
In this way, a unique assignment of number states is not possible since the $B$-species shows mainly a delocalized behavior 
over both wells. 
For completeness we should mention that the dynamics in region II is described by a higher-order superposition 
involving all available number states i.e. $\ket{3,0}_B$, $\ket{2,1}_B$, $\ket{0,3}_B$, $\ket{1,2}_B$. 
Entering region III, we can observe that the motion of the $B$-species alternates during the time evolution. 
Indeed, at the initial stages of the dynamics ($10<t<50$) all three $B$-species fermions tunnel to the right-hand well, see in 
particular that $p^{(2)}_{BB}(10<t<50)\approx1$ and $\braket{ \rho^{(1)}_B(10<t<50)}_L\approx0$. 
For later time instants the cloud predominantly oscillates between $\ket{2,1}_B$ (e.g. $p^{(2)}_{BB}(t\approx60)\approx0.4$ and $\braket{ \rho^{(1)}_B(t\approx60)}_L\approx2$)   
and $\ket{0,3}_B$ (e.g. $p^{(2)}_{BB}(t\approx100)\approx1$ and $\braket{ \rho^{(1)}_B(t\approx100)}_L\approx0$). 
In this way we can infer that the fermions in this regime of interactions tunnel as pairs \cite{chen2011controlling,meinert2014f,mistakidis2017mode,mistakidis2014interaction}.   

To understand further the correlated dynamics within the different tunneling regions we additionally analyze  
the interspecies pair probability $p^{(2)}_{AB}(t)$ presented in Fig. \ref{abb:explefth1} (e) and Table \ref{table}. 
Indeed within the region I $p^{(2)}_{AB}(t)$ undergoes small amplitude oscillations, see 
e.g. $p^{(2)}_{AB}(t\approx30)\approx2/3$ and $p^{(2)}_{AB}(t\approx60)\approx0.85$. 
The latter indicate that here the most significant number state configurations of the system 
correspond to the $\ket{0,3}_A\otimes \ket{1,2}_B$ and $\ket{3,0}_A\otimes\ket{2,1}_B+\ket{3,0}_A\otimes\ket{3,0}_B$ respectively. 
Turning to region II, $0.2<p^{(2)}_{AB}(t)<0.7$ and therefore a clear assignment in terms of number states can not be unavoided. 
However this suggests, as it has already been observed in $p^{(2)}_{AA}(t)$ and $p^{(2)}_{BB}(t)$, the participation of a multitude 
of number states and as a result the occurrence of a multimode dynamics. 
Is is also worth stressing again that this interaction regime II exhibits a strongly alternating behavior with respect to $g_{AB}$ , 
e.g. see $\braket{ \rho^{(1)}_{\sigma}(t)}_L$, $p^{(2)}_{\sigma\sigma'}(t)$ for $0.2<g_{AB}<1.7$ and $1.7<g_{AB}<3.0$ in Fig. \ref{abb:explefth1}. 
Moreover, as we shall argue below the dynamics in this region is characterized by an enhanced degree of both entanglement and fragmentation, see Sec. \ref{entanglement}.  
Finally, for strong interactions (region III) $p^{(2)}_{AB}(t)$ shows a much more pronounced oscillatory pattern when compared to regions I and II. 
Since here the $A$-species is fully trapped in the left well this behavior of $p^{(2)}_{AB}(t)$ is attributed to the motion of the $B$-species 
which perform pair tunneling, i.e. $\ket{2,1}_B\leftrightarrow\ket{0,3}_B$.

\subsection{Entanglement and Fragmentation} \label{entanglement}

To expose the many-body nature of the dynamics of the FF mixture we next measure the degree of entanglement (or interspecies correlations) 
and the fragmentation (or intraspecies correlations) of the underlying many-body state.   
These features can be quantified by employing $S(t)$ and $F_{\sigma}(t)$ respectively for varying $g_{AB}$, see Fig. \ref{abb:entanglement} and Table \ref{table}. 
We remind here that when $S(t)=0$ the system is called non-entangled, while $S(t)\neq0$ signifies the presence of entanglement 
or interspecies correlations. 
Accordingly, for $F_{\sigma}(t)>0$ the $\sigma$-species is termed intraspecies correlated (see also Sec. \ref{observables}).  
Most importantly, since the fermions of each species are spin-polarized, i.e. non-interacting, the occurrence of intraspecies 
correlations during the evolution is induced by the interspecies correlations. 
Recall also that in the initial (ground) state of the system the distinct fermionic clouds that reside in the 
left well, see Fig. \ref{abb:sketch} (a), are non-overlaping and 
therefore fully phase separated due to their mass imbalance \cite{shin2006observation,shin2008phase}.  
Due to this non-overlaping behavior no inter- and intraspecies correlations occur for arbitrary $g_{AB}$. 
As it can be easily seen in Figs. \ref{abb:entanglement} (a), (b), (c) for weak interactions (region I) there is only a small 
amount of inter- and intraspecies correlations since both $S(t)$ and $F_{\sigma}(t)$ are suppressed. 
Strikingly enough, for intermediate interactions (region II) $S(t)$ as well as $F_{\sigma}(t)$ increase during 
the evolution and tend to saturate to a certain finite value, 
signifying that the many-body state is strongly entangled and fragmented. 
Inspecting region III we observe again that both $S(t)$ and $F_{\sigma}(t)$ increase in the course of time but 
in a slower manner as compared to region II. 
Also they acquire moderate magnitudes with respect to region II. 
Interestingly enough within this strongly interacting regime III, $F_{B}(t)$ (heavier species) increases more rapidly 
during the evolution when compared to $F_{A}(t)$ (lighter species). 

\begin{figure}
	\includegraphics[width=0.5\textwidth]{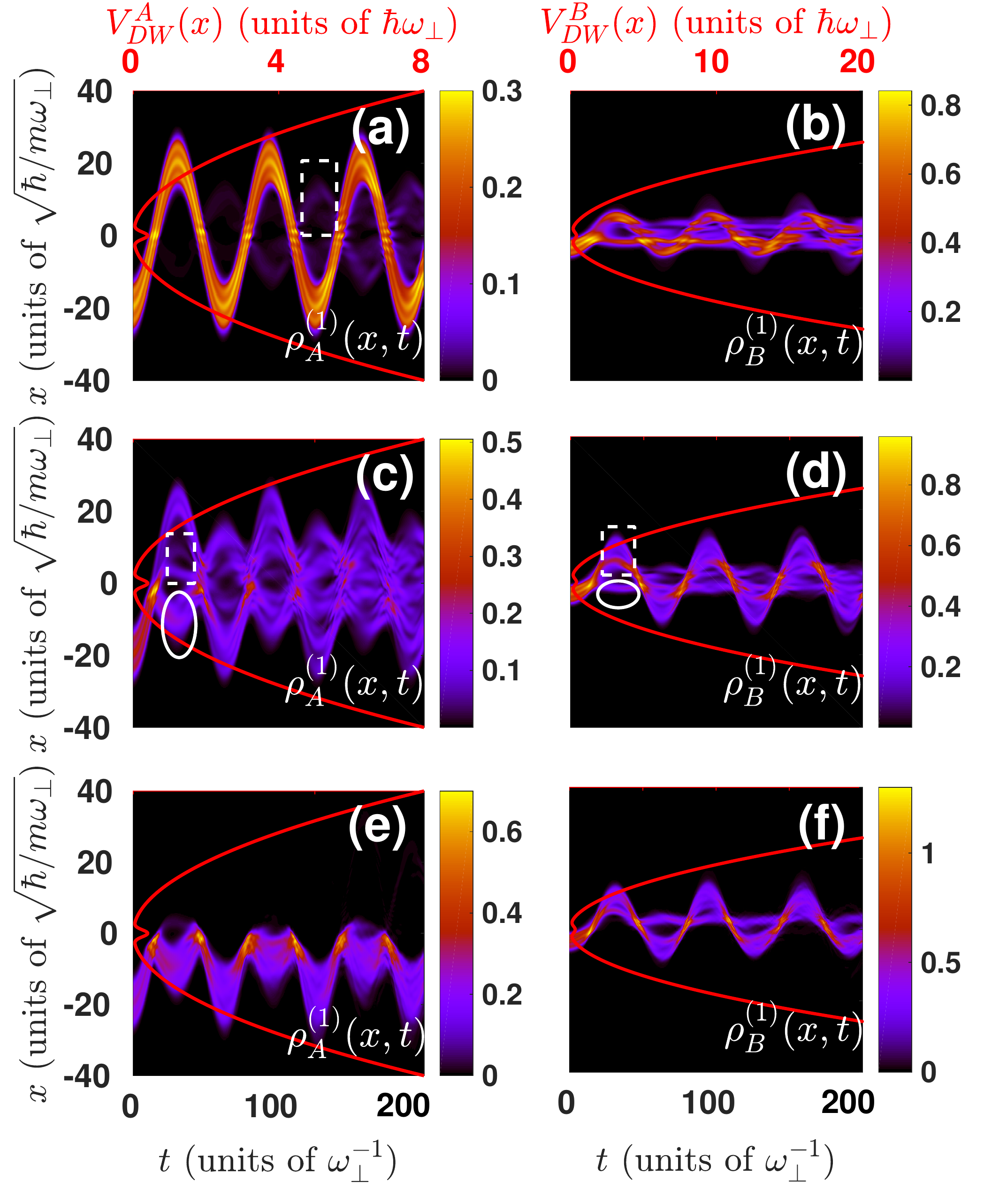}
	\caption{ Evolution of the single-particle density for the $A$ (left column) and the $B$-species (right column) of a FF mixture 
	consisting of $N_A=N_B=3$ fermions after a quench of the tilt parameter from $d=0.2$ to $d=0$. 
	The interspecies interaction corresponds to (a), (b) $g_{AB}=0.1$, (c), (d) $g_{AB}=1.2$ and (e), (f) $g_{AB}=4$ respectively. 
	The dashed rectangles and solid ellipses indicate the fermionic fragments that tunnel and are reflected back respectively 
	due to the barrier or a collision with the other species. 
	The double-well that each species experiences, $V_{DW}^{\sigma}$, is also shown on top of the density evolution.} 
	\label{abb:1bdensityh1} 
\end{figure}

\subsection{Single-Particle Density Evolution}\label{single_particle_density} 

To visualize the tunneling motion in the different interaction regimes identified above, we finally invoke 
the evolution of the single-particle density, $\rho^{(1)}_{\sigma}(t)$, for each of the species see Fig. \ref{abb:1bdensityh1}. 
Due to its larger mass the $B$-species resides near the barrier, while the lighter $A$-species is in comparison shifted to the left. 
Within the region I, see Figs. \ref{abb:1bdensityh1} (a) and (b), the $A$-species cloud oscillates as a whole back and forth 
resulting in a three-body tunneling process. 
Only for large evolution times, $t>100$, a small fraction of the $A$-species gets reflected from the barrier, see the dashed 
rectangle in Fig. \ref{abb:1bdensityh1} (a). 
For the $B$-species a substantial reflection already occurs during the first tunneling process and a corresponding fraction remains 
trapped in the left well. 
Consequently a delocalization of $\rho^{(1)}_{B}(t)$ takes place. 
At intermediate interspecies interactions (region II) both species (and especially the lighter $A$-species) are reflected from 
both the barrier as well as from the other species, see Figs. \ref{abb:1bdensityh1} (c), (d). 
In particular, during the initial stages of the dynamics ($0<t<25$) both the $A$- and the $B$-species travel towards the barrier 
where they split into two fragments from which one tunnels through the barrier to the right well 
[see the dashed rectangles in Figs. \ref{abb:1bdensityh1} (c), (d)] and the other one is reflected back to the 
left well [see the solid ellipses in Figs. \ref{abb:1bdensityh1} (c), (d)]. 
Then, the transmitted fragments of the $A$- and the $B$-species collide in the right well [see e.g. the dashed rectangle in Fig. \ref{abb:1bdensityh1} (c)]. 
As a result a further split into two new fragments takes place. 
One of these fragments travels back to the barrier and the other one continues towards the outer part of the right well. 
The effect of this collisional dynamics is, of course, much more pronounced for the $A$-species since it is the lighter one \cite{ML-MCTDHX} and it is 
manifested by the delocalization of both species over the wells. 
This delocalization becomes even more apparent for large evolution times, $t>100$, and it is a consequence of the fact that the distinct species 
suffer multiple collisions with one another, rendering the one-body density blurred. 
Recall also that the dynamics in this regime of interactions is strongly entangled and fragmented, see Fig. \ref{abb:entanglement}. 
Let us also comment at this point that within the Hartree-Fock approximation the above-mentioned delocalization of both species during the evolution 
becomes suppressed (not shown here for brevity) since it is essentially caused by the contribution of the higher-lying orbitals. 
Turning to the strongly interacting regime (region III), we observe that the heavier $B$-species acts as a hard-wall or material barrier 
for the lighter $A$-species \cite{pflanzer2009material,pflanzer2010interspecies}, as shown in Figs. \ref{abb:1bdensityh1} (e), (f). 
In this way, the $A$-species remains fully trapped within the left well during the evolution, while the $B$-species still 
performs a tunneling motion. 
The only effect of the collision between the two species that is imprinted in the density of the $B$-species is a splitting of its density 
into two parts, one of which undergoes tunneling between the two wells and the other remains trapped in the right well where it performs 
tiny amplitude oscillations. 
\begin{figure}[t]
	\includegraphics[width=0.5\textwidth, height=0.4\textheight]{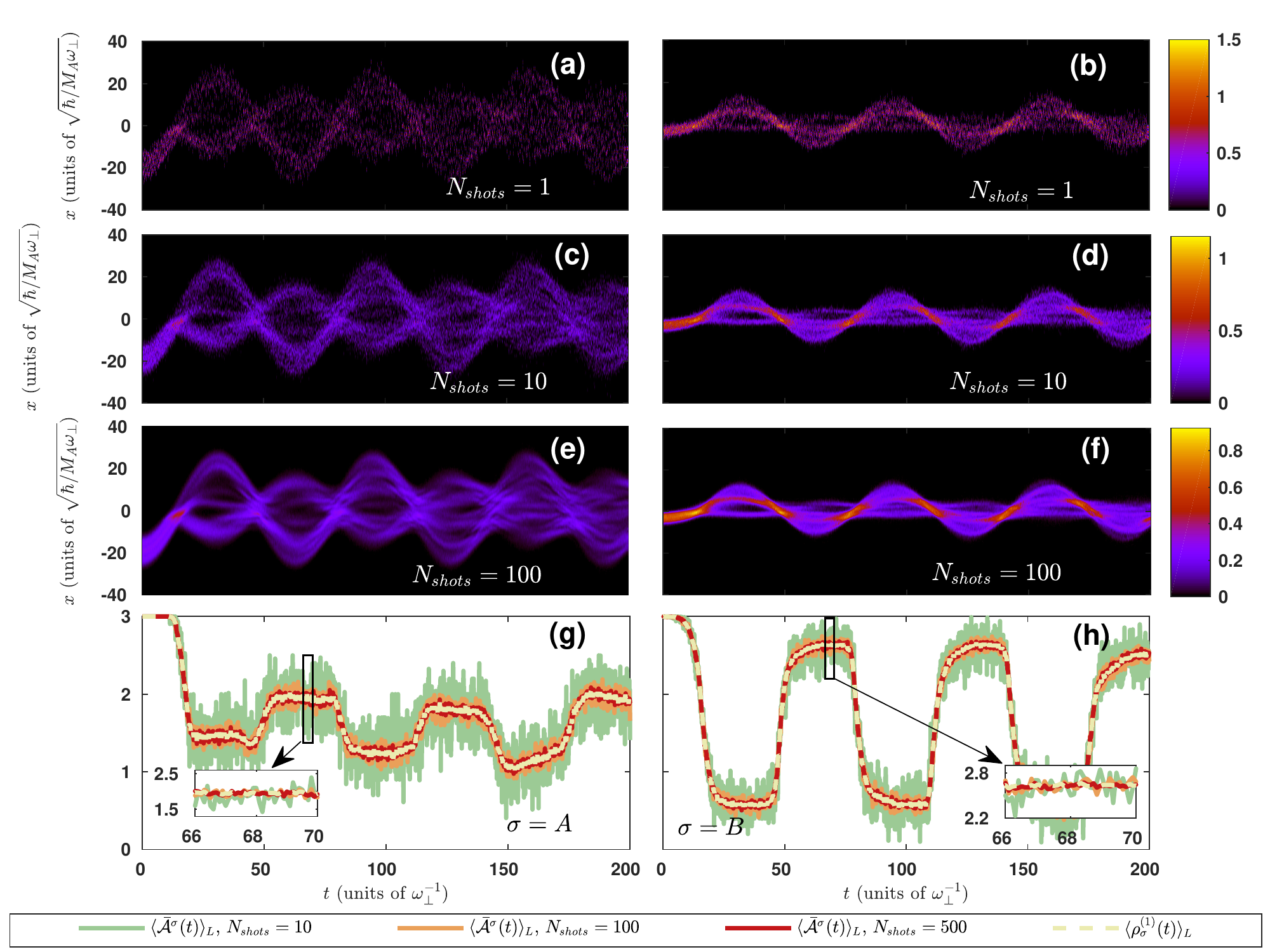}
	\caption{ Averaged images over (a), (b) $N_{shots}=1$, (c), (d) $N_{shots}=10$, 
	(e), (f) $N_{shots}=100$.  
	(g), (h) Evolution of $\braket{ \rho^{(1)}_{\sigma}(t)}_L$ and the probabilities, $\braket{\bar{\mathcal{A}}^{\sigma}(t)}$, 
	obtained by averaging several single-shot images (see legend). 
	The insets illustrate $\braket{\bar{\mathcal{A}}^{\sigma}(t)}$ in certain time intervals. 
	In all cases the system consists of $N_A=N_B=3$ fermions with $M_B/M_A=6$ and $g_{AB}=1.2$, while the dynamics is 
	induced by performing a quench of the initially tilted double-well ($d=0.2$) to a symmetric one with $d=0$. 
	Left panels indicate the $A$-species and right panels refer to the $B$-species.}
	\label{abb:shots} 
\end{figure}

\section{Single-Shot Simulations}\label{shots}

To provide possible experimental evidence for the many-body tunneling dynamics of the FF mixture 
we simulate in-situ single-shot absorption measurements \cite{sakmann2016single,mistakidis2017correlation,lode2017fragmented}. 
This type of measurements probe the spatial configuration of the atoms which can be inferred by the many-body 
probability distribution. 
Such experimental images refer to a convolution of the spatial particle configuration with a 
point spread function which essentially dictates the corresponding experimental resolution. 
For our calculations, to be presented below, we use a point spread function of Gaussian shape possessing a 
width $w_{PSF}=1 \ll l\approx 3.2$, where $l=\sqrt{1/\omega}$ denotes the corresponding harmonic oscillator length. 
It is also important to note here that in few-body experiments \cite{zurn2012fermionization,wenz2013few} besides the in-situ imaging of the cloud 
another technique to probe the state of the system and get rid of unavoidable noise sources that might 
destroy the experimental signal is fluorescence imaging \cite{serwane2011deterministic,koutentakis2018probing}.  
However, the simulation of this technique which has been proven important for few-body experiments lies 
beyond our current scope. 
Here we aim at demonstrating how in-situ imaging can be used to adequately monitor the above-described few-body 
FF mixture dynamics. 
\begin{figure*}[t]
	\includegraphics[width=\textwidth]{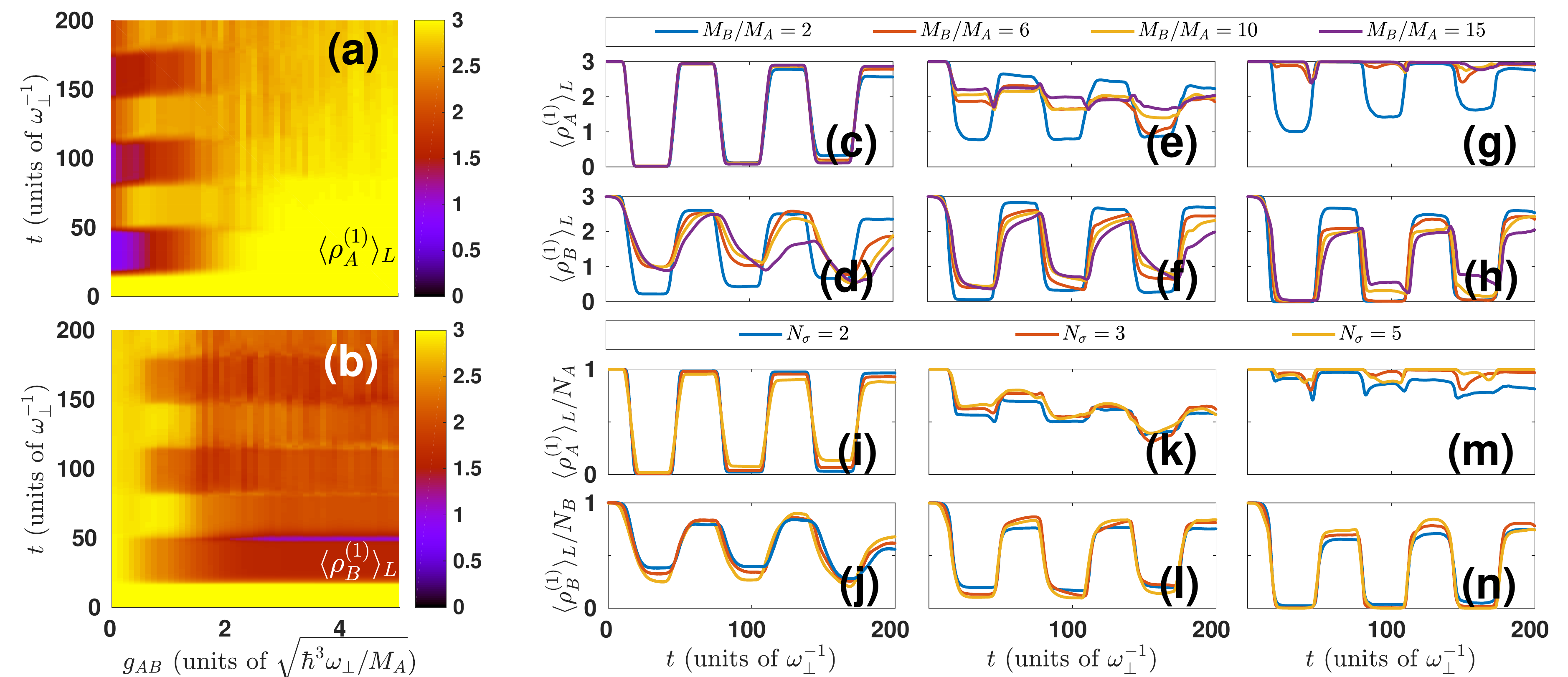}
	\caption{ Evolution of $\braket{\rho^{(1)}_{\sigma}(t)}_L$, of the (a) $\sigma=A$ and (b) $\sigma=B$ species with mass ratio $M_B/M_A=6$ 
	residing in a double-well with barrier height $V_0=4$, for increasing interspecies repulsion $g_{AB}$.
	(c)-(h) $\braket{\rho^{(1)}_{\sigma}(t)}_L$ for different mass ratios (see legend) and interspecies repulsion 
	(c), (d) $g_{AB}=0.1$, (e), (f) $g_{AB}=1.5$ and (g), (h) $g_{AB}=4.0$. 
	(a)-(h) The system consists of $N_A=N_B=3$ fermions. 
	(i)-(n) $\braket{\rho^{(1)}_{\sigma}(t)}_L/N_{\sigma}$ for varying number of fermions (see legend) with $M_B/M_A=6$ and interspecies repulsion 
	(i), (j) $g_{AB}=0.1$, (k), (l) $g_{AB}=1.5$ and (m), (n) $g_{AB}=4.0$. 
	The corresponding barrier height of the double-well in (c)-(n) is $V_0=1$. 
	In all cases the FF mixture is initially confined in a tilted double-well with $d=0.2$ and 
	the dynamics is induced by quenching to a symmetric double-well i.e. $d=0$.}
	\label{abb:dependency} 
\end{figure*}

Relying on the system's many-body wavefunction obtained within ML-MCTDHX we simulate in-situ single shot images for both 
species $A$, $\mathcal{A}^A(\tilde{x};t)$, and species $B$, $\mathcal{A}^B(\tilde{x'}|\mathcal{A}^A(\tilde{x});t_{im})$, 
at each instant $t$ of the evolution when we consecutively image first the $A$ and then the $B$ species. 
For details regarding the simulation process of this experimental technique we refer to Appendix B. 
Below we focus on the FF dynamics, possessing $g_{AB}=1.2$ (region II), within the double-well upon quenching 
the tilt parameter from $d=0.2$ to $d=0$. 
Note that a similar procedure has been followed also for other values of $g_{AB}$ (not shown here). 
Figures \ref{abb:shots} ($a$), ($b$) show the first simulated in-situ single-shot images for both species during the evolution, 
namely $\mathcal{A}^A(\tilde{x};t)$, and $\mathcal{A}^B(\tilde{x'}|\mathcal{A}^A(\tilde{x});t)$ respectively. 
It can be deduced that the two species exhibit a tunneling behavior, resembling this way the overall tendency 
observed in the one-body density [see also Figs. \ref{abb:1bdensityh1} (c), (d)]. 
However, it is important to mention that a direct correspondence between the one-body density and one single-shot image 
is not possible due to the small particle number of the considered FF mixture, $N_A=N_B=3$. 
Another source of the inability to explicitly observe the one-body density within a single-shot image is the 
presence of multiple orbitals in the system. 
Indeed, the many-body state builds upon a superposition of multiple orbitals [see Eqs. (\ref{Eq:WF}) and (\ref{Eq:SPFs})] 
and therefore imaging an atom alters the many-body state of the other atoms and hence their one-body density. 
For a more elaborated discussion on this topic see \cite{mistakidis2017correlation,katsimiga2017many,katsimiga2018many}. 
To retrieve the one-body density of the system we next rely on an average of several single-shot images for each species, 
namely $\bar{\mathcal{A}}^A(\tilde{x};t)=1/N_{shots}\sum_{k=1}^{N_{shots}} 
\mathcal{A}_k^A(\tilde{x};t)$ and $\bar{\mathcal{A}}^B(\tilde{x}^{'}|\mathcal{A}^A(\tilde{x});t)=1/N_{shots}\sum_{k=1}^{N_{shots}} 
\mathcal{A}_k^B(\tilde{x}^{'}|\mathcal{A}^A(\tilde{x});t)$ respectively. 
In particular, Figs. \ref{abb:shots} (c)-(f) illustrate $\bar{\mathcal{A}}^A(\tilde{x};t)$ and 
$\bar{\mathcal{A}}^B(\tilde{x}^{'}|\mathcal{A}^A(\tilde{x});t)$ for different number of samplings i.e. $N_{shots}$. 
Evidently, a comparison of this averaging process for an increasing number of shots and the actual one-body density 
obtained within ML-MCTDHX [see Figs. \ref{abb:1bdensityh1} (c), (d)] reveals that they are almost identical. 
Namely as the number of shots, $N_{shots}$, becomes larger then $\bar{\mathcal{A}}^A(\tilde{x};t)$ and $\bar{\mathcal{A}}^B(\tilde{x}^{'}|\mathcal{A}^A(\tilde{x});t)$ 
tend gradually to $\rho^{(1)}_{A}(t)$ and $\rho^{(1)}_{B}(t)$ respectively. 
To further support our above-mentioned arguments we finally contrast the one-body tunneling probabilities $\braket{ \rho^{(1)}_A(t)}_L$ and 
$\braket{ \rho^{(1)}_B(t)}_L$ at $g_{AB}=1.2$ with the corresponfing simulated and experimentally to be observed probabilities i.e. 
$\braket{\bar{\mathcal{A}}^A(t)}=\int_{\tilde{x}<0} d\tilde{x}\bar{\mathcal{A}}^A(\tilde{x};t)$ and 
$\braket{\bar{\mathcal{A}}^B(t)}=\int_{\tilde{x}^{'}<0}d\tilde{x}^{'}\bar{\mathcal{A}}^B(\tilde{x}^{'}|\mathcal{A}^A(\tilde{x});t)$ 
[see Figs. \ref{abb:shots} (g), (h)]. 
As it can be readily seen, a larger number of $N_{shots}$, see in particular the insets in Figs. \ref{abb:shots} (g), (h), (here $N_{shots}>100$) reproduces almost perfectly 
both $\braket{ \rho^{(1)}_A(t)}_L$ and $\braket{ \rho^{(1)}_B(t)}_L$, capturing this way the tunneling process.

\section{Further Characteristics of the Tunneling Dynamics}\label{control}
 
Having discussed in detail the properties of the tunneling dynamics of the FF mixture in the double-well, we 
finally showcase how the overall dynamics depends on certain system parameters. 
To this end, we first study the effect of the barrier height $V_0$ on the quench induced tunneling dynamics. 
As a reference system we consider the double-well with $V_0=1$ that has already been discussed above, while as an indicator 
for the tunneling dynamics we employ $\braket{\rho^{(1)}_{\sigma}(t)}_L$. 
Figures \ref{abb:dependency} (a), (b) present $\braket{\rho^{(1)}_{A}(t)}_L$ and $\braket{\rho^{(1)}_{B}(t)}_L$ respectively for the case $V_0=4$ for 
varying interspecies repulsion $g_{AB}$. 
We observe that the tunneling dynamics of both species, and especially the $B$-species, is strongly influenced by the barrier height and in particular it is 
overall suppressed \cite{mistakidis2015negative,sowinski2016diffusion,fasshauer2016multiconfigurational}, compare Figs. \ref{abb:explefth1} (a), (b) and Figs. \ref{abb:dependency} (a), (b). 
Regarding the $A$-species the three above-identified tunneling regions are shifted to weaker interactions 
and in particular region I becomes negligible in size and exists only very close to the non-interacting limit [hardly visible in Fig. \ref{abb:dependency} (a)]. 
Therefore, an increasing barrier height shows a similar effect on the $A$-species as a stronger interaction strength 
where the $B$-species acts as a material barrier \cite{pflanzer2009material,pflanzer2010interspecies}, see for instance Fig. \ref{abb:1bdensityh1} (e). 
For the heavier $B$-species the barrier height has the most prominent effect as the tunneling dynamics is significantly suppressed 
for every value of $g_{AB}$ when compared to the $V_0=1$ case where tunneling oscillations occur independently of the interaction strength. 
Here the tunneling takes place for $g_{AB}>1$ and it exhibits a small amplitude.  

As a next step, we examine the influence of the mass ratio on the tunneling behavior of the FF mixture 
by inspecting again $\braket{\rho^{(1)}_{\sigma}(t)}_L$ within the different interaction regimes. 
Since a mass ratio $M_B/M_A>15$ leads to a ground state where both species are not fully trapped in the left well 
for an initial tilt magnitude $d=0.2$, we investigate below only imbalances that satisfy $M_B/M_A\in [2,15]$. 
In particular we show $\braket{\rho^{(1)}_{\sigma}(t)}_L$ in each of the regions I, II and III for 
$M_B/M_A\in \{2,6,10,15\}$, see Figs. \ref{abb:dependency} (c)-(h). 
Referring to weak interactions, see Figs. \ref{abb:dependency} (c), (d), we observe that an increasing mass ratio mainly affects 
the heavier $B$-species. 
Indeed, the amplitude of $\braket{\rho^{(1)}_{B}(t)}_L$ is reduced for a larger $M_B$, accompanied by an overall damping 
in the course of time. 
In contrast $\braket{\rho^{(1)}_{A}(t)}_L$ is essentially independent of the increasing $M_B$ with the only noticeable difference being a slight 
lowering of the damping amplitude of the $\braket{\rho^{(1)}_{A}(t)}_L$ oscillation for sufficiently large evolution times. 
Turning to intermediate interactions (region II) we deduce that the oscillation amplitude of $\braket{\rho^{(1)}_{\sigma}(t)}_L$ decreases during evolution while its damping  
increases for both species for increasing mass ratios, see Figs. \ref{abb:dependency} (e), (f). 
This damping behavior is much stronger for the lighter $A$-species as it be seen by comparing Figs. \ref{abb:dependency} (e) and (f). 
Interestingly enough, an irregular behavior of $\braket{\rho^{(1)}_{B}(t)}_L$ takes place at the turning points of the tunneling motion 
with a tendency to a sawtooth waveform. 
For strong interspecies interactions (region III) the tunneling dynamics of the $A$-species vanishes for larger $M_B$. 
However, the $B$-species exhibits prominent tunneling oscillations which show a damping behavior for increasing $M_B$. 
Indeed, according to our observations for strong $g_{AB}$ in the case of $M_B/M_A=6$ [Figs. \ref{abb:1bdensityh1} (e), (f)] 
the heavier $B$-species acts as an effective material barrier \cite{pflanzer2009material,pflanzer2010interspecies} for the 
lighter $A$-species and its collision with the $B$-species pushes the latter to the right well, enforcing its tunneling motion. 

To shed light on the particle number dependence of the tunneling motion, we finally utilize the normalized expectation value 
$\langle\rho_\sigma^{(1)}(t)\rangle_L/N_\sigma$ for a varying particle number 
$N_\sigma \in \{2,3,5\}$ within the different interaction regimes, see Figs. \ref{abb:dependency} (i)-(n). 
Overall, we observe that for all three interaction regimes (I, II, III) an increasing particle number results in a stronger  
damping of $\langle\rho_\sigma^{(1)}(t)\rangle_L/N_{\sigma}$ for the lighter $A$-species, while the corresponding damping 
of the heavier $B$-species is washed out. 
All the other one-body tunneling features, e.g. the frequency and the amplitude of 
the $\langle\rho_\sigma^{(1)}(t)\rangle_L/N_{\sigma}$ oscillation, remain essentially insensitive. 
An inspection of the two-body tunneling characteristics unveils more pronounced differences since for larger 
particle numbers more modes are triggered which are imprinted e.g. in the $p^{(2)}_{\sigma \sigma ^\prime}$ evolution (results not shown here).

\section{Conclusions}\label{conclusions}

We have investigated the tunneling dynamics of a FF mixture with spin polarized components confined in a double-well. 
The fermionic mixture is initially prepared within the left well of a tilted double-well and the tunneling dynamics 
is induced by removing the tilt and let the system evolve in a symmetric double-well. 
We particularly examined the impact of the interspecies interactions on the tunneling behavior of each species and unveiled 
the significant role of intra- and interspecies correlations. 
The emergent dynamics has been characterized on both the one- and two-body level by invoking the single- and two-particle 
probabilities for the fermions of each species to reside e.g. in the left-hand well in the course of the dynamics. 

For very weak interactions close to the non-interacting limit both species perform almost perfect Rabi-oscillations.  
The heavier species exhibit small deviations from the perfect Rabi scenario since they either show single or two-particle tunneling 
i.e. during the evolution not all three particles tunnel together. 
Increasing the interspecies repulsion the tunneling dynamics is suppressed for both species. 
The lighter species undergoes single-particle tunneling, while for the heavier one a more complex dynamics takes place which is 
characterized by a higher-order quantum superposition further indicating the presence of strong correlations in the system. 
Turning to strong interactions we observe that the lighter species experiences a quantum self-trapping due to the heavier species which 
acts as a material barrier. 
In particular, the collision between the two species causes a splitting of the density of the heavier component into two parts. 
The first part performs tunneling between the two wells and the other one remains trapped in the right well exhibiting 
small amplitude oscillations. 
As a consequence, this heavier component undergoes almost perfect Rabi-oscillations being characterized either by three-particle or pair 
tunneling at different instants of the evolution. 
The degree of both inter- and intraspecies correlations for all interaction regimes has been exposed and found to be overall
significant especially for intermediate interactions. 
To relate our findings with possible experimental realizations we simulate in-situ single-shot measurements and showcase how an increasing 
sampling of such images can be used to adequately reproduce the observed fermionic tunneling dynamics. 

The dependence of the observed tunneling behavior on the mass ratio of the two components, the particle number 
and the height of the barrier of the double-well has been discussed. 
It is shown that the mass imbalance between the components possesses a strong influence on the tunneling dynamics 
depending on the interspecies interaction strength. 
Namely, for weaker interactions the tunneling amplitude of the heavier species becomes smaller for an increasing mass ratio, while 
the tunneling features of the lighter species remain essentially unaffected. 
Increasing the repulsion a decrease of the tunneling amplitude for both species occurs during evolution. 
For strong interactions the heavier species acts as a material barrier for the lighter one thus suppressing the tunneling motion of the latter. 
On the other hand, for fixed interspecies repulsion and larger particle numbers a damping of the tunneling oscillations takes place especially 
for the lighter species. 
Finally, we show that for a higher barrier the tunneling dynamics of both species, and in particular of the heavier one, is overall suppressed 
since an increasing barrier disfavors the tunneling process. 

There is a multitude of interesting directions worth pursuing in future studies. 
A straightforward one would be to examine the tunneling dynamics of a dipolar FF mixture, under the quench 
protocol considered herein, and investigate how the long-range character of the interactions alters the 
emergent tunneling behavior including the inherent entanglement generation. 
Another interesting prospect is to unravel the collisional dynamics of two fermi components which are initially well 
separated, e.g. due to the presence of a high barrier and then are left to collide by removing this intermediate 
barrier in a sense of the counterflow experiment. 
Inspecting the many-body character of the spontaneously generated non-linear excitations such as dark solitons \cite{scott2011dynamics} 
or domain-wall structures \cite{katsimiga2017many,katsimiga2018many} between the two species is a challenging future task. 
Finally, the generalization of our current findings utilizing time-dependent quench protocols would be desirable 
in order to design schemes for selective transport between the wells of each component.

\appendix

\section{Tunneling Dynamics Versus Energy Offset} \label{sec:tilt_magnitude} 

Let us briefly comment on the influence of the quench strength, i.e. the value of the postquench tilt parameter $d$, on the tunneling dynamics 
of both species of the FF mixture. 
We consider a FF mixture of $N_A=N_B=3$ with $M_B/M_A=6$ and $g_{AB}=0.1$ trapped initially in a tilted double-well with $d=0.2$ and $h=1$. 
To induce the tunneling dynamics we perform a quench to a smaller value of the tilt $d\in [0,0.2)$, thus rendering the energy offset between the wells smaller and favoring 
the tunneling motion of both species from the left to the right well.

\begin{figure}
	\includegraphics[width=0.5\textwidth]{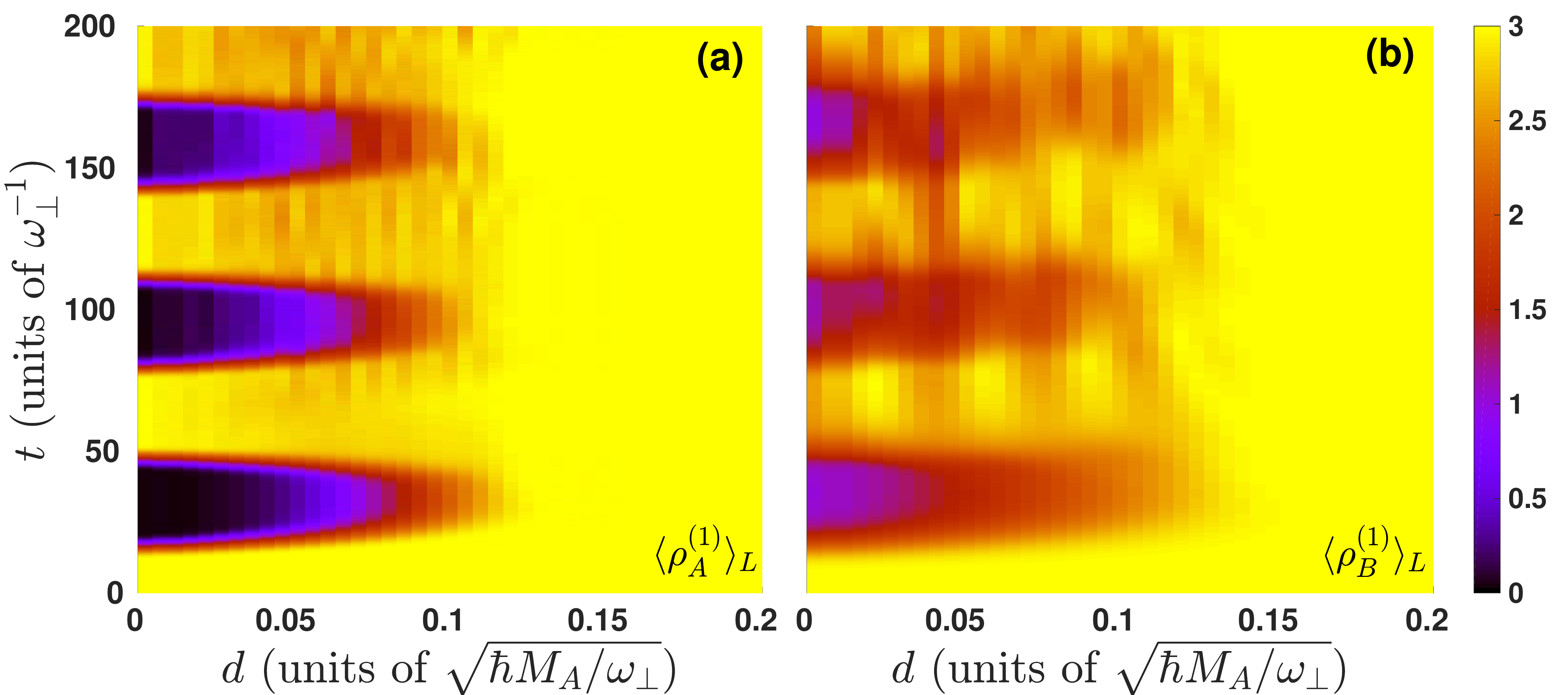}
	\caption{ $\braket{\rho^{(1)}_{\sigma}(t)}_L$, of the (a) $\sigma=A$ and (b) $\sigma=B$ species 
	for varying postquench tilt parameter $d$. 
	The mixture consists of $N_A=N_B=3$ fermions repulsively interacting with $g_{AB}=0.1$ and $M_B/M_A=6$ and being initially 
	trapped in a tilted double-well with $d=0.2$ and $V_0=1$. 
	The dynamics is induced by a quench on the tilt parameter $d$.}
	\label{abb:tiltdep} 
\end{figure} 

Figures \ref{abb:tiltdep} (a), (b) present $\braket{\rho^{(1)}_{\sigma}(t)}_L$ for a varying postquench amplitude of the tilt magnitude $d$. 
For offsets characterized by $0<d<0.03$ we do not observe any significant influence on the tunneling oscillations of both species. 
Entering the region $0.03<d<0.09$ significant alterations in $\braket{\rho^{(1)}_{\sigma}(t)}_L$ occur in the sense that the oscillation frequency 
and amplitude decrease for a larger offset. 
Finally, for $d>0.09$ both species remain in the left well without tunneling throughout the evolution 
as the energy offset is adequately large and suppresses any possible tunneling process. 
We remark that we have performed the same investigation for both intermediate and strong interspecies interactions $g_{AB}$, observing 
a similar behavior of the emerging dynamics within the same postquench intervals of the tilt magnitude $d$ as above (not shown here for brevity).

\section{Single-Shots Implementation in Fermi-Fermi Mixtures} \label{single_shots}

To simulate the experimental single-shot procedure we perform a sampling of 
the many-body probability distribution \cite{sakmann2016single,katsimiga2017many,katsimiga2018many,mistakidis2017correlation} 
being accessible within the ML-MCTDHX framework. 
We remark that such a theoretical implementation of the experimental process has already been performed 
for single-species bosons \cite{sakmann2016single,lode2017fragmented,katsimiga2017many,katsimiga2018many} 
and fermions \cite{koutentakis2018probing} as well as for binary bosonic ensembles \cite{mistakidis2017correlation} but 
not yet for FF mixtures. 
As in the two-species bosonic case, the corresponding single-shot procedure for FF mixtures 
depends strongly on the inter- and intraspecies correlations inherent in the system. 
Indeed within a many-body treatment the presence of entanglement between the different species 
is of significant importance regarding the image ordering. 
This can be understood by inspecting the Schmidt decomposition [see Eq. (\ref{Eq:WF})] and especially the 
involved Schmidt coefficients $\lambda_k$'s. 
Below let us elaborate on the corresponding process where we image first the $A$ and then the $B$-species, obtaining 
in this way the images $\mathcal{A}^A(\tilde{x})$ and $\mathcal{A}^B(\tilde{x'}|\mathcal{A}^A(\tilde{x}))$. 
However, we note that the same overall procedure has to be followed in order to image first the $B$ and then the $A$ species, 
resulting in single-shots images $\mathcal{A}^B(\tilde{x})$ and 
$\mathcal{A}^A(\tilde{x'}|\mathcal{A}^B(\tilde{x}))$ respectively. 

To image first the $A$ and then the $B$ species we consecutively annihilate all the $N_A$ fermions. 
Referring to a specific time instant of the imaging process, e.g. $t_{im}$, a random position 
is drawn that satisfies the constraint $\rho_{N_A}^{(1)}(x_1')>q_1$, where $q_1$ denotes a random 
number lying in the interval [$0$, $ \max\lbrace{\rho^{(1)}_{N_A}(x;t_{im})\rbrace}$]. 
Next, we project the ($N_A+N_B$)-body wavefunction onto the ($N_A-1+N_B$)-body one. 
The latter is accomplished by the use of the projection operator $\frac{1}{\mathcal{N}}(\hat{\Psi}_A(x_1')\otimes \hat{\mathbb{I}}_B)$, where 
$\hat{\Psi}_A(x_1')$ refers to the fermionic field operator that annihilates an $A$ species fermion located at $x_1'$ and $\mathcal{N}$ 
is the normalization constant. 
An important observation, here, is that the above process affects the Schmidt weights, $\lambda_k$, and therefore 
besides that the $B$-species has not already been imaged, both $\rho^{(1)}_{N_A-1}(t_{im})$ and $\rho^{(1)}_{N_B}
(t_{im})$ are altered. 
To comprehend the latter we rely on the Schmidt decomposition according to which the many-body wavefunction 
after the first measurement reads
\begin{equation}
\begin{split}
&\ket{\tilde{\Psi}_{MB}^{N_A-1,N_B}(t_{im})}=\\ &\sum_i \sqrt{\tilde{\lambda}_{i,N_A-1}(t_{im})}\ket{\tilde{\Psi}_{i,N_A-1}^A(t_{im})}\ket{\Psi_i^B(t_{im})}.   
\label{Eq:A1}
\end{split}
\end{equation} 
In this expression, $\ket{\tilde{\Psi}_{i,N_A-1}^A}=\frac{1}{N_i}\hat{\Psi}_A(x_1')\ket{\Psi_i^A}$ denotes the 
$N_A-1$ species wavefunction, $N_i=\sqrt{\bra{\Psi_i^A}\hat{\Psi}_A^{\dagger}(x_1')\hat{\Psi}_A(x_1')\ket{\Psi_i^A}}$ is 
the normalization factor and $\tilde{\lambda}_{i,N_A-1}=\lambda_i N_i/\sum_i \lambda_i N_i^2$ refer to the Schmidt 
coefficients of the ($N_A-1+N_B$)-body wavefunction. 
Repeating the above steps $N_A-1$ times we obtain the following distribution of positions ($x'_1$, $x'_2$,...,$x'_{N_A-1}$) 
which is then convoluted with a point spread function. 
This results in the single-shot image $\mathcal{A}^A(\tilde{x})=\sum_{i=1}^{N_A}e^{-\frac{(\tilde{x}-x'_i)^2}{2w_{PSF}^2}}$ 
for the $A$-species, where $\tilde{x}$ are the spatial coordinates within the image and $w_{PSF}$ is the width of the employed 
point spread function. 
It is also important to mention at this point that after annihilating all $A$-species fermions the many-body 
wavefunction reads 
\begin{equation}
\begin{split}
&\ket{\tilde{\Psi}_{MB}^{0,N_B}(t_{im})}=\\ &\ket{0} \otimes\sum_i \frac{\sqrt{\tilde{\lambda}_{i,1}(t_{im})}
\braket{x|\Phi_{i,1}^A}}{\sum_j{\sqrt{\tilde{\lambda}_{j,1}(t_{im})|\braket{x|\Phi_{j,1}^A}|^2}}}\ket{\Psi_i^B(t_{im})},   
\label{Eq:A3}
\end{split}
\end{equation} 
where $\braket{x|\Phi_{j,1}^A}$ refers to the single-particle orbital of the $j$-th mode and 
the wavefunction of the $B$-species, $\ket{\Psi_{MB}^{N_B}(t_{im})}$, corresponds to the second 
term of the cross product on the right-hand side. 
Evidently, $\ket{\Psi_{MB}^{N_B}(t_{im})}$ obtained after the annihilation of all $N_A$ fermions corresponds to a 
non-entangled $N_B$-particle many-body wavefunction and its corresponding single-shot procedure reduces to that of the 
single-species case \cite{sakmann2016single,katsimiga2017many,katsimiga2018many}. 
The latter has been extensively used in a variety of settings (see e.g. \cite{sakmann2016single,katsimiga2017many,katsimiga2018many}) and therefore 
it is only briefly outlined below. 
For a time instant of the imaging $t=t_{im}$ we calculate $\rho^{(1)}_{N_B}(x;t_{im})$ 
from the many-body wavefunction $\ket{\Psi_{N_B}}\equiv \ket{\Psi(t_{im})}$. 
Next, a random position $x''_1$ is drawn according to the constraint $\rho^{(1)}_{N_B}(x''_1;t_{im})>q_2$, 
with $q_2$ being a random number within [$0$, $\rho^{(1)}_{N_B}(x;t_{im})$]. 
Then, one particle at position $x''_1$ is annihilated and the $\rho^{(1)}_{N_B-1}(x;t_{im})$ is calculated 
from $\ket{\Psi_{N_B-1}}$. 
To proceed, a new random position $x''_2$ is drawn from $\rho^{(1)}_{N_B-1}(x;t_{im})$. 
We repeat this procedure for $N_B-1$ steps and obtain the distribution of positions 
($x''_1$, $x''_2$,...,$x''_{N_B-1}$) which is then convoluted with a point spread function  
leading to a single-shot image $\mathcal{A}^B(\tilde{x'}|\mathcal{A}^A(\tilde{x}))$.

\section{Remarks on Convergence of the Many-Body Simulations} \label{sec:numerics}

Let us finally briefly discuss the basic aspects of our numerical method and then showcase the convergence of our results. 
ML-MCTDHX \cite{ML-MCTDHX} consists of a variational method for solving the time-dependent 
many-body Schr{\"o}dinger equation of atomic mixtures with constituents being 
either of Bose \cite{mistakidis2017correlation,katsimiga2017dark} or Fermi \cite{cao2017collective,koutentakis2018probing} type. 
To construct the many-body wavefunction a time-dependent variationally optimized many-body basis is employed, enabling us to take into account 
the system's important correlation effects using a computationally feasible basis size. 
Therefore the system's relevant subspace of the Hilbert space is spanned at each time instant of the evolution in a more efficient manner 
when compared to expansions relying on a time-independent basis as the number of basis states can be significantly reduced. 
Finally, as a result of the multi-layer ansatz for the total wavefunction we are able to capture both the intra- and the interspecies 
correlations emerging during the non-equilibrium dynamics of a bipartite system.

\begin{figure}
	\includegraphics[width=0.48\textwidth]{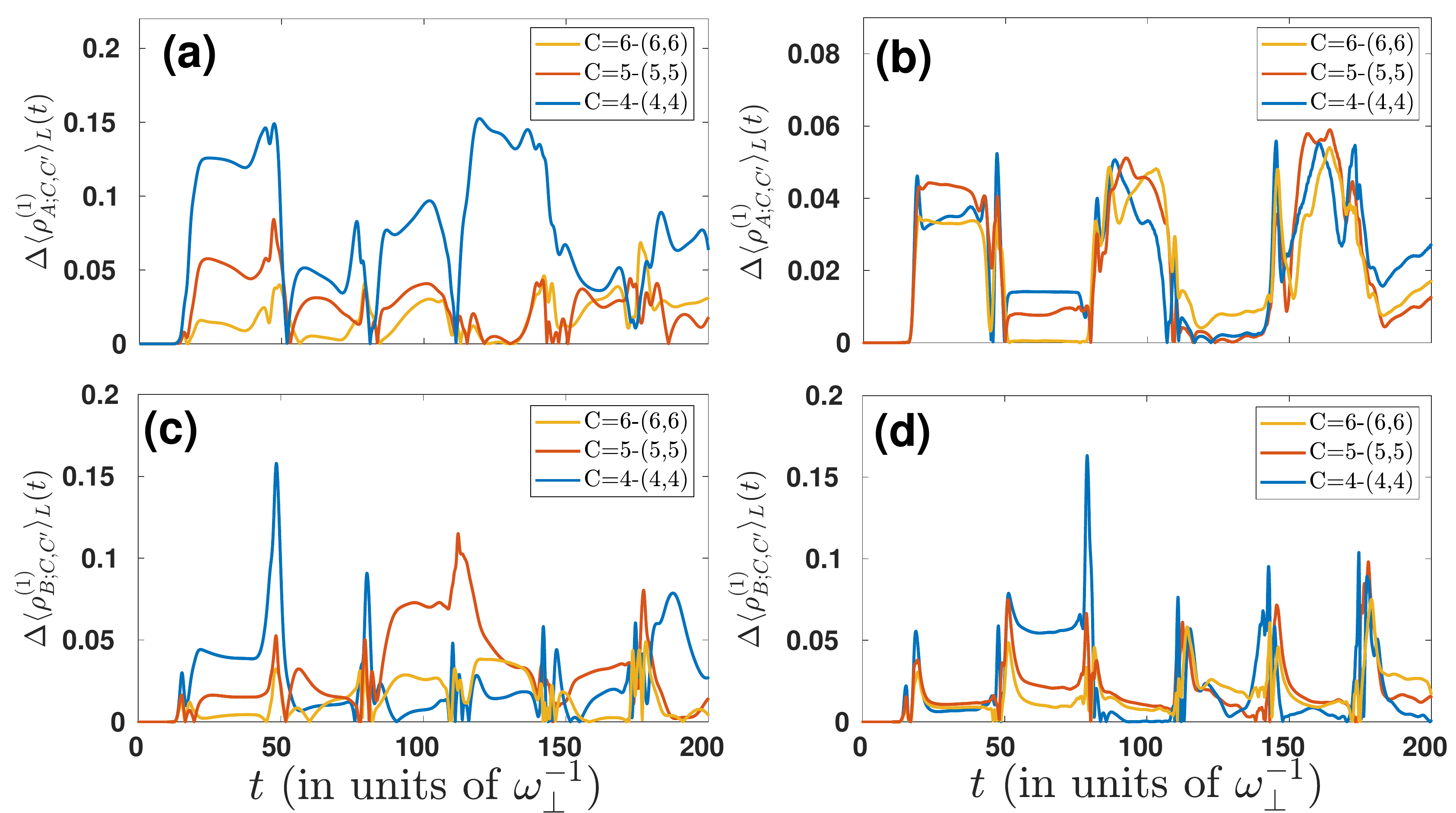}
	\caption{Evolution of the deviations $\Delta\langle\rho^{(1)}_{\sigma;C,C^\prime}(t)\rangle_L$ between the $C^\prime=9-(9,9)$ 
	and other numerical configurations $C=M-(m^A,m^B)$ (see legends) for (a), (c) $g_{AB}=1.2$ and (b), (d): $g_{AB}=4.0$. 
	The particle number of each component is $N_A=N_B=3$. }
	\label{abb:convergence} 
\end{figure}

Within our approach the used Hilbert space truncation, namely the order of the considered approximation, 
corresponds to the employed numerical configuration space being denoted by $C=M-(m^A,m^B)$. 
In this notation $M=M^A=M^B$ and $m^A$, $m^B$ refer to the number of species and single-particle 
functions respectively for each of the species. 
We also note that for our simulations, a primitive basis consisting of 
a sine discrete variable representation is employed which involves 400 grid points. 
To infer about the convergence of our many-body simulations we systematically check that upon varying 
the numerical configuration space $C=M-(m^A,m^B)$ the observables of interest remain insensitive. 
We remark that all many-body calculations discussed in the main text rely on the numerical 
configuration $C=9-(9,9)$ for $N_A=N_B=3$ and $C=10-(10;10)$ when $N_A=N_B=5$. 
In this way, the available Hilbert space for the simulation includes 8037 (9140) coefficients 
when $N_{\sigma}=3$ ($N_{\sigma}=5$). 
This is in sharp contrast to an exact diagonalization procedure which should take into account 10586800 (83219 $10^{10}$) 
coefficients for the $N_{\sigma}=3$ ($N_{\sigma}=5$) case, rendering these simulations infeasible. 
To conclude, let us briefly showcase the convergence of our results upon varying the number of 
species functions and single-particle functions. 
To perform this investigation we employ the expectation value of the one-body density in the left well, $\langle \rho^{(1)}_{\sigma;C}(t)\rangle_L$, 
and calculate its absolute deviation during the time evolution, for each of the species, between the $C'=(9;9;9)$ and other  
numerical configurations $C=M-(m^A,m^B)$  
\begin{equation}
	\Delta\langle\rho^{(1)}_{\sigma;C,C^\prime}(t)\rangle_L =\frac{|\langle \rho^{(1)}_{\sigma;C}(t)\rangle_L-\langle \rho^{(1)}_{\sigma;C^\prime}(t)\rangle_L|}{N_\sigma}, \label{convergence} 
\end{equation}
where $N_{\sigma}$ denotes the particle number of the $\sigma$-species. 
As it is evident from Eq. (\ref{convergence}) $\Delta\langle\rho^{(1)}_{\sigma;C,C^\prime}(t)\rangle_L$ is normalized to unity. 
Fig. \ref{abb:convergence} ($a$) [($b$)] shows $\Delta\langle\rho^{(1)}_{A;C,C^\prime}(t)\rangle_L $ [$\Delta\langle\rho^{(1)}_{B;C,C^\prime}(t)\rangle_L $] 
following a quench of the tilt parameter from $d=0.2$ to $d=0$ for $g_{AB}=1.2$. 
For reasons of completeness we remark that this value of $g_{AB}$ lies within the region that the 
tunneling dynamics is characterized by strong inter- and intraspecies correlations, see also Fig. \ref{abb:entanglement}. 
We indeed observe an adequate convergence of both $\Delta\langle\rho^{(1)}_{A;C,C^\prime}(t)\rangle_L $ and 
$\Delta\langle\rho^{(1)}_{B;C,C^\prime}(t)\rangle_L $.  
In particular, comparing the $C=6-(6,6)$ and $C'=9-(9,9)$ approximations, the corresponding relative difference, for both species, 
is less than $5\%$ throughout the evolution. 
The same observations can also be made for the case of strong interspecies interactions as 
illustrated in Figs. \ref{abb:convergence} (c) and (d) for species $A$ and $B$ respectively. 
Finally, we remark that a similar analysis has been performed for all other observables used in the main text and found to be adequately converged, i.e. 
their relative deviations between the $C'=9-(9,9)$ and the $C=6-(6,6)$ configurations lie below $6\%$ (not shown here for brevity).

\section*{Acknowledgements} 
S.I.M. and P.S. gratefully acknowledge financial support by the Deutsche Forschungsgemeinschaft 
(DFG) in the framework of the SFB 925 ``Light induced dynamics and control of correlated quantum
systems''.

\bibliography{bibliography}
\end{document}